\documentclass[a4paper]{article}

\usepackage{a4wide}
\usepackage{amsmath,amssymb,amsfonts,amsthm,latexsym}
\usepackage{xcolor,color}
\usepackage{graphicx}
\usepackage{datetime}
\usepackage[pdftex,
breaklinks=true,bookmarks=true,colorlinks=true,linkcolor=blue,citecolor=blue,allcolors=blue]{hyperref}
\usepackage[position=top]{subfig}
\usepackage{algorithm}
\usepackage{algpseudocode}
\usepackage[sort&compress,authoryear,round]{natbib}
\usepackage{doi}
\usepackage{relsize}
\usepackage{filecontents}
\usepackage{enumerate}
\usepackage{lmodern}
\usepackage{xspace}

\hypersetup{
    unicode=false,          pdftoolbar=true,        pdfmenubar=true,        pdffitwindow=false,     pdfstartview={FitH},    pdfcreator={Creator},   pdfkeywords={keyword1, key2, key3}, pdfnewwindow=true,      colorlinks=true,        linkcolor=blue,         citecolor=blue,         filecolor=magenta,      urlcolor=cyan           }

\usepackage{cleveref}
    \crefname{section}{Section}{Sections}
    \Crefname{section}{Section}{Sections}
    \crefname{subsection}{Section}{Sections}
    \Crefname{subsection}{Section}{Sections}
    \crefname{figure}{Fig.}{Figs.}
    \Crefname{figure}{Figure}{Figures}
    \crefname{equation}{Eq.}{Eqs.}
    \Crefname{equation}{Equation}{Equations}
    \crefname{table}{Tab.}{Tabs.}
    \Crefname{table}{Table}{Tables}
    \crefname{appendix}{Appendix}{Appendices}
    \Crefname{appendix}{d}{Appendices}
    \setcitestyle{citesep={,}}

\usepackage{tikz}
\usetikzlibrary{decorations.pathreplacing}
\usetikzlibrary{external,shapes}

\usepackage{soul}

\definecolor{myorange}{rgb}{0.93, 0.69, 0.13}
\definecolor{mygray}{rgb}{0.5, 0.5, 0.5}

\definecolor{cS04}{rgb}{0.000000, 0.000000, 1.000000}
\definecolor{cS06}{rgb}{1.000000, 0.000000, 0.000000}
\definecolor{cS08}{rgb}{0.000000, 1.000000, 0.000000}
\definecolor{cS10}{rgb}{0.000000, 0.000000, 0.172414}
\definecolor{cS12}{rgb}{1.000000, 0.103448, 0.724138}
\definecolor{cS16}{rgb}{1.000000, 0.827586, 0.000000}
\definecolor{cS20}{rgb}{0.000000, 0.344828, 0.000000}
\definecolor{cS25}{rgb}{0.517241, 0.517241, 1.000000}
\definecolor{cS30}{rgb}{0.620690, 0.310345, 0.275862}
\definecolor{cS35}{rgb}{0.000000, 1.000000, 0.758621}
\definecolor{cS40}{rgb}{0.000000, 0.517241, 0.586207}
\definecolor{cS60}{rgb}{0.000000, 0.000000, 0.482759}
\definecolor{cS80}{rgb}{0.586207, 0.827586, 0.310345}

\DeclareRobustCommand\straightline{\tikzset{external/export=false}\tikz[baseline=-0.6ex]{\draw[-,line width=1mm] (-1ex,.5ex)--++(.3cm,0);
\tikzset{external/export=true}
}~}

\newcommand{\blackline}{
\tikzset{external/export=false}\tikz[baseline]{
\draw[-,line width=.25mm,black] (-1ex,.5ex)--++(.3cm,0);
\tikzset{external/export=true}
}~}

\newcommand{\blueline}{
\tikzset{external/export=false}\tikz[baseline]{
\draw[-,line width=.25mm,blue] (0,.5ex)--++(.3cm,0);
\tikzset{external/export=true}
}~}

\newcommand{\blackdashed}{
\tikzset{external/export=false}\tikz[baseline]{
\draw[-,line width=.25mm,dashed] (0,.5ex)--++(.5cm,0);
\tikzset{external/export=true}
}~}

\newcommand{\oblacksolid}{
\tikzset{external/export=false}\tikz[baseline]{
\node[draw,scale=0.5,circle,line width=.25mm] () at (.25cm,.5ex){};
\draw[-,line width=.25mm,solid] (0,.5ex)--++(.5cm,0);
}~}

\newcommand{\oblue}{
\tikzset{external/export=false}\tikz[baseline]{
\node[draw=blue,scale=0.5,circle,line width=.25mm] () at (.25cm,.5ex){};
\tikzset{external/export=true}
}~}

\newcommand{\sblacksolid}{
\tikzset{external/export=false}\tikz[baseline]{
\node[draw,scale=0.5,regular polygon, regular polygon sides=4,line width=.25mm]
() at (.25cm,.5ex){};
\draw[-,line width=.25mm,solid] (0,.5ex)--++(.5cm,0);
\tikzset{external/export=true}
}~}

\newcommand{\tblacksolid}{
\tikzset{external/export=false}\tikz[baseline]{
\node[draw,scale=0.35,regular polygon, regular polygon sides=3,line width=.25mm]
() at (.25cm,.5ex){};
\draw[-,line width=.25mm,solid] (0,.5ex)--++(.5cm,0);
\tikzset{external/export=true}
}~}

 \renewcommand{\v}[1]{\ensuremath{\mathbf{#1}}} \newcommand{\gv}[1]{\ensuremath{\boldsymbol{#1}}}

\newcommand{\mean}[1]{\langle #1 \rangle}

 \renewcommand{\d}[2]{\dfrac{\mbox{d}{#1}}{\mbox{d}{#2}}}
 \newcommand{\pd}[2]{\frac{\partial #1}{\partial #2}}

\newcommand{\pdinline}[2]{{\partial{#1}}/{\partial{#2}}}

\newcommand{\diff}{\mathop{}\!\mathrm{d}}

\newcommand{\umean}{\mean{u_f}}

\newcommand{\urms}{\langle u_f^\prime u_f^\prime\rangle_{xz}^{1/2}}
\newcommand{\vrms}{\langle v_f^\prime v_f^\prime\rangle_{xz}^{1/2}}
\newcommand{\wrms}{\langle w_f^\prime w_f^\prime\rangle_{xz}^{1/2}}

\newcommand{\vort}{\gv{\omega}_{f}}

\newcommand{\kenfrisch}{K}
\newcommand{\ken}{\kenfrisch}
\newcommand{\kenspec}{\ken^{ps}}
\newcommand{\kenfd}{\ken^{fd}}

\newcommand{\reynolds}{Re}
\newcommand{\wllx}{\lambda_x}

\newcommand{\wllz}{\lambda_z}

\newcommand{\rhof}{\rho_f}
\newcommand{\rhop}{\rho_p}
\newcommand{\uf}{\v{u}_f}
\newcommand{\up}{\v{u}_p}
\newcommand{\omegap}{\gv{\omega}_p}

\newcommand{\hydrostress}{\gv{\tau}}
\newcommand{\gravity}{\v{g}}
\newcommand{\domain}[1]{\mathcal{L}_{#1}}
\newcommand{\dpart}{D_{p}}
\newcommand{\dplus}{\dpart^+}
\newcommand{\tscalep}{\tau_p}

\newcommand{\tscaleb}{\tau_b}
\newcommand{\tscalenu}{\tau_\nu}

\newcommand{\galileo}{Ga}
\newcommand{\stokes}{St}
\newcommand{\stokesplus}{\stokes^+}

\newcommand{\dissipation}{\mathcal{E}_D}
\newcommand{\utau}{u_\tau}

\newcommand{\deltanu}{\delta_\nu}
\newcommand{\retau}{Re_\tau}

\newcommand{\wallstu}{\tau_{w,u}}
\newcommand{\wallstl}{\tau_{w,l}}

\newcommand{\zpar}{z_p}
\newcommand{\upar}{u_p}

\begin{document}
\epstopdfsetup{suffix=}
\author{
{T. Pestana}\footnote{t.pestana@tudelft.nl}\\\small Aerodynamics Group, Faculty of Aerospace Engineering\\
\small Technische Universiteit Delft\\
\small Kluyverweg 2, 2629 HS Delft, The Netherlands\\[3ex]
Markus Uhlmann\footnote{markus.uhlmann@kit.edu}\\
\small Insititute for Hydromechanics,\\
\small Karlsruhe Insititute of Technology, \\
\small Karlsruhe, 76131, Germany\\[3ex]
Genta Kawahara\footnote{kawahara@me.es.osaka-u.ac.jp}\\
\small Graduate School of Engineering Science, \\
\small Osaka University, 1-3 Machikaneyama, Toyonaka, \\
\small Osaka 560-8531, Japan
}

\title{Can preferential concentration of finite-size particles in plane
    Couette turbulence be reproduced with the aid of equilibrium solutions?
}
\date{\small \today\\[1ex] \small (accepted for publication in \textit{Phys.
Rev. Fluids} (2020))}

\maketitle

\begin{abstract}
This work employs for the first time invariant solutions of the Navier-Stokes
equations to study the interaction between finite-size particles and near-wall
coherent structures. We consider horizontal plane Couette flow and focus on
Nagata's upper-branch equilibrium solution \citep{Nagata1990a} at low Reynolds
numbers where this solution is linearly stable. When adding a single heavy
particle with a diameter equivalent to 2.5 wall units (one twelfth of the gap
width), we observe that the solution remains stable and is essentially unchanged
away from the particle. This result demonstrates that it is technically feasible
to utilize exact coherent structures in conjunction with particle-resolved DNS.
While translating in the streamwise direction, the particle migrates laterally
under the action of the quasi-streamwise vortices until it reaches the region
occupied by the low-speed streak, where it attains a periodic state of motion --
independent of its initial position. As a result of the ensuing preferential
particle location, the time-average streamwise particle velocity differs from
the plane-average fluid-phase velocity at the same wall-distance as the particle
center, as previously observed in experiments and in numerical data for fully
turbulent wall-bounded flows. Additional constrained simulations where the
particle is maintained at a fixed spanwise position while freely translating in
the other two directions reveal the existence of two equilibria located in the
low-speed and in the high-speed streak, respectively, the former being an
unstable point. A parametric study with different particle to fluid density
ratios is conducted which shows how inertia affects the spanwise fluctuations of
the periodic particle motion. Finally, we discuss a number of potential future
investigations of solid particle dynamics which can be conducted with the aid of
invariant solutions (exact coherent structures) of the Navier-Stokes equations.
\end{abstract}

\section{Introduction}
In many natural and technical systems the interaction between fluid flow and
solid particles plays an important role. Examples are the most diverse: in
nature it ranges from blood flow to volcanic eruptions, whereas in industrial
applications a classical example is the combustion of pulverized coal. In such
systems, one aspect that is of primordial importance is the spatial distribution
of the dispersed phase. It can significantly affect various quantities of
practical interest, such as particle dispersion, inter-particle collision
statistics, mean relative velocities and turbulence enhancement/attenuation.
Nevertheless, to the present day, our understanding of the fundamental process
behind structure formation in the particulate phase or about turbulence
modulation due to particles is still incomplete.

A typical approach to understand fluid-particle interaction consists of
analyzing data from direct numerical simulations (DNS). These analyses are often
performed \textit{a posteriori} and consist in operations like spatial filtering
or conditional averaging in an attempt to extract the key mechanisms that govern
fluid-particle interaction. Approaches of this kind have led to new observations
and have advanced the knowledge on this topic quite considerably in the past
decades. For example, in wall-bounded flows, heavy particles moving
near a solid
wall tend to concentrate in the low-speed regions and form streamwise aligned
streaks. This particle preferential concentration has been observed in numerous
experimental and numerical investigations, (e.g.,
Refs.~\citep{yung:1989,rashidi:1990a,hetsroni:1994,kaftori:1995a,nino:1996,pan:1997,uhlmann:2008,manolo:2012}), and it is known to
depend on particle size and density. \citet{kidanemariam:2013} performed
particle-resolved DNS of horizontal turbulent channel flow with a very dilute
set of particles having a diameter equivalent to 7 wall units and a submerged
weight which essentially restricted them to a position in contact with the lower
bounding wall. Through conditional averaging they found that there is a
significant statistical correlation between spanwise particle velocity and the
presence of a nearby quasi-streamwise vortex with the corresponding sign of
rotation. This statistical argument explains the migration to low-speed streaks
which are typically flanked by counter-rotating streamwise vortices; as a
consequence, the average streamwise particle velocity is lower than the mean
fluid velocity at the same wall-distance due to the preferential particle
concentration in low-speed regions. Despite the success of such statistical
approaches it is not always straightforward to extract the salient features of a
fully turbulent flow field from massive DNS data-sets, and to link the relevant
flow structures to the observed particle dynamics.

An alternative to the above methods is to consider surrogate flow-fields instead
of fully-developed turbulence. This has the obvious advantage of making the
analysis of particle motion conceptually easier, although at the risk of being
less relevant to the fully turbulent state of interest. Along this line,
\citet{maxey:2002} used a cellular flow to investigate the clustering properties
of small spherical particles, and \citet{reeks:2006} considered randomized
Taylor-Green vortex flow to elucidate the random uncorrelated particle motion.
Of equal interest is the work of \citet{bergougnoux:2014}, who set up
experimentally a cellular flow to investigate the influence of vortical
structures upon the settling of small particles.

Another way to simplify the problem, which has to the best of our knowledge not
yet been explored, is to investigate particle motion in flow fields that are
invariant solutions to the Navier-Stokes equations. These solutions are widely
believed to be of relevance to full turbulence, as some of them are able to
reproduce statistical features of turbulent flows; cf.\ \citet{Kawahara2011a}
for a review about the significance of invariant solutions to turbulence, and
\citet{vanveen:2019} for a review on the history of invariant solutions in
turbulence. Invariant solutions range from simple fixed points (equilibrium
states), which are time invariant in some suitable frame of reference, to more
complex and dynamic flow fields that recur at a fixed time period (periodic
orbits). Solutions of this type have been found in various flow configurations,
such as plane Couette flow \citep{Nagata1990a}, plane Poiseuille flow
\citep{waleffe:2001}, Hagen-Poiseuille flow \citep{Faisst2003}, or square-duct
flow \citep{uhlmann:10a,okino:10}. In the present contribution we focus on plane
Couette flow, which on one hand is a wall-bounded flow with far-reaching
applications, and for which, on the other hand, well-known solutions exist.

Equilibrium states for plane Couette flow have traditionally been found through
homotopy techniques, i.e.\ by embedding the problem of interest into a
generalized configuration: \citet{Nagata1990a} used spanwise rotation,
\citet{Clever1992} resorted to wall heating, and \citet{Waleffe1998} imposed an
artificial body force. Nagata's solution originates from a saddle-node
bifurcation at a finite Reynolds number, and it can be continued along a lower
and an upper solution branch, both of which extend to high-Reynolds numbers. The
upper branch of the solution is believed to be of relevance for the turbulent
state: the coherent structures embedded therein, i.e.\ a pair of
counter-rotating streamwise vortices flanking a wavy low-speed streak, emulate
well the coherent structures typically found in the turbulent buffer layer, and
the velocity profiles of these solutions also resemble those typically found in
turbulent wall-bounded flows \citep{Jimenez2005a}. Let us mention in passing
that periodic orbits containing a full regeneration cycle of vortices and
velocity streaks have been detected in plane Couette flow by
\citet{Kawahara2001a} and others.

Thus, some of the known invariant solutions, more specifically equilibrium
solutions, provide exact coherent structures that are of relevance to
turbulence, like an apparatus in a laboratory that would contain some of the
main ingredients of turbulence \citep{kerswell:05}, albeit typically at low
Reynolds number. This unique trait is appealing for the study of the effects of
coherent vortices on particle dynamics, as it makes the problem more amenable by
untangling turbulence and isolating its building blocks. However, this approach
is technically not trivial as the presence of a solid phase can disturb the
equilibrium state and lead to re-laminarization or transition to turbulence. In
a first step towards introducing the concepts of invariant solutions into the
study of finite-size particle motion, the present work considers the question
whether simple invariant solutions seeded with finite-size particles can be used
to study the interaction between finite-size particles and true turbulence. To
this end, we address the following specific questions in the context of plane
Couette flow:
\begin{enumerate}[(i)]
\item  How long can equilibrium solutions be observed once finite-size particles
have been added?
\item Do equilibrium solutions seeded with finite-size particles exhibit
    preferential particle concentration?
\end{enumerate}
In \cref{sec:methodology} we describe the flow configuration, present the
numerical methodology, and define the physical and numerical parameters of this
study. In the sequence (\cref{sec:trackinvsol}), we compute a set of equilibrium
solutions and study their stability. Finally, in \cref{sec:invsolparticles}, we
add finite-size particles to these plane Couette flow equilibria and study the
ensuing temporal evolution of the system.

\section{Numerical Set-up}\label{sec:methodology}

\subsection{Flow configuration and parameters}

\begin{figure}[p]
\begin{center}
\includegraphics[scale=1]{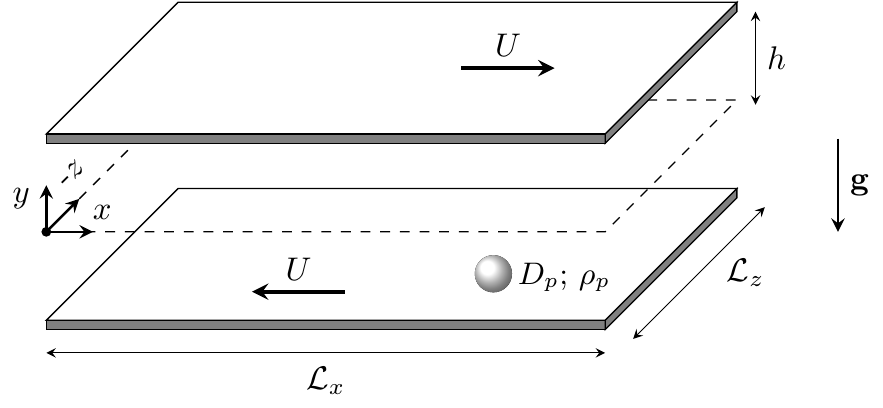}
\caption{Plane Couette flow geometry illustrating the numerical set-up. The
physical domain is assumed periodic in the streamwise $(x)$ and spanwise
directions $(z)$, whereas the $y$-direction is bounded by two walls that move in
opposite directions, each with speed $U$. When a solid-phase is present, it
consists of finite-size spherical particles with diameter $\dpart$ and density
$\rhop$. Gravity $\mathbf{g}$ acts in the negative $y$-direction.}
\label{fig:couette_geometry}
\end{center}
\end{figure}

We consider plane Couette flow with two parallel walls, which are separated by a
distance $2h$, and extend in the $x$ and $z$ directions by $\domain{x}$ and
$\domain{z}$, such that the physical domain is defined as
$\gv{\Omega}=[0,\domain{x}]\times[-h,h]\times[0,\domain{z}]$ (cf.\
\cref{fig:couette_geometry}). The walls move in the $x$-direction with speed
$U$, and their counter movement induces shear and drives the flow. A
gravitational field $\v{g}$ acts in the negative wall normal direction with
intensity $g$.

The fluid motion is governed by the incompressible Navier-Stokes equations, viz.
\begin{align}
  \label{eq:ns_mass}
  & \nabla \cdot \uf = 0, \\[1ex]
& \pd{\uf}{t} + (\uf \cdot \nabla)\, \uf
    = -\frac{1}{\rhof} \nabla p + \nu
  \nabla^2 \uf,
  \label{eq:ns_momentum}
\end{align}
where, $\uf=(u_f,v_f,w_f)$ is the fluid velocity, $p$ is the hydrodynamic
pressure, $\nu$ is the kinematic viscosity of the fluid and $t$ denotes time.
The flow field $\uf$ satisfies a no-slip boundary condition at the two solid
walls and at the fluid/particle interface. The flow field is assumed periodic in
the $x$ and $z$ directions with streamwise and spanwise wavelengths
$\lambda_x=\domain{x}$ and $\lambda_z=\domain{z}$, respectively.

The motion of the solid-phase, which is formed of finite-size spherical
particles of diameter $\dpart$ and density $\rhop$, is governed by the
Newton-Euler equations for the motion of rigid bodies:
\begin{align}
\label{eq:nl_linearmom}
& \rhop V_p \d{\up}{t} = \rhof \oint_{\partial S} \hydrostress \cdot \v{n} \,
\mathrm{d}\sigma + (\rhop - \rhof)V_p \gravity, \\[1ex]
& I_p \d{\omegap}{t} = \rhof \oint_{\partial S} \v{r} \times (\hydrostress \cdot
\v{n}) \mathrm{d}\sigma.
\label{eq:nl_angularmom}
\end{align}
Here $\up=(u_p,v_p,w_p)$ is the linear particle velocity, $\omegap$ is the
angular particle velocity, $\gv{\tau}=-p\,\v{I} + 2\nu(\v{u}_f+\v{u}_f^{T})$ is
the hydrodynamic stress with $\v{I}$ the identity tensor and $\v{r}$ the
position vector with respect to the particle's centroid.
The vector normal to the fluid-solid interface $\partial S$ is denoted by
$\v{n}$, and the volume and the moment of inertia of the particles are $V_p$ and
$I_p$, respectively. At the fluid/particle interface $\partial S$, a no-slip
boundary condition is imposed through an immersed boundary method as further
explained in \cref{sec:fdsolver} and in \cref{apx:ibm}. Particle-wall contact
is assumed to be frictionless, as in the DNS of \citet{kidanemariam:2013}.
Particles can, therefore, slide without additional tangential stress along the
horizontal walls.

In its most general form, our numerical experiments involve a total of $9$
control parameters that need to be prescribed:
$\{\rhop,\rhof,\nu,\domain{x},\domain{z},h,\dpart,U,g\}$. This specific set may
be combined into a group of $6$ independent non-dimensional numbers. These are
taken as the dimensionless geometrical dimensions $\domain{x}/h$ and
$\domain{z}/h$, the Reynolds number based on the absolute wall speed and on half
of the wall separation $\reynolds=Uh/\nu$, the density ratio $\rhop/\rhof$, the
non-dimensional particle diameter $\dpart/h$ and the Galileo number
$Ga=u_g\dpart/\nu$, where $u_g=\sqrt{(\rhop/\rhof-1) g \dpart}$ is the
gravitational velocity.

The box-averaged energy dissipation rate $\dissipation$, is defined as
\begin{equation}
  \dissipation =
  \frac{ h^2 } {U^2} \mean{\vort\cdot\vort}_{xyz},
  \label{eq:dissipation}
\end{equation}
where $\vort$ denotes the fluid vorticity, and $\mean{\,\cdot\,}_{xyz}$ is the
spatial average over the three spatial directions. Note that the dissipation
defined in (\cref{eq:dissipation}) is normalized such that the laminar flow
solution yields $\dissipation=1$. The instantaneous wall shear stress, averaged
over each of the two wall planes, is defined as:
\begin{align}
    \wallstl=\rhof\nu\,\pd{\umean_{xz}}{y}\Bigr|_{y=-h}
    \quad \text{and} \quad
    \wallstu=\rhof\nu\,\pd{\umean_{xz}}{y}\Bigr|_{y=+h}
\end{align}
where $\mean{\,\cdot\,}_{xz}$ denotes the two-dimensional spatial averaging
operator.
By defining the mean friction velocity as $\utau=\sqrt{(\wallstl +
\wallstu)/(2\rhof)}$, the viscous lengthscale is given by $\deltanu=\nu/\utau$,
and the friction-velocity-based Reynolds number is $\retau=\utau h/\nu$.
Henceforth the standard notation with a ``$+$'' superscript refers to quantities
normalized with these viscous wall units. The bulk velocity is defined as
\begin{equation}
  u_b = \frac{1}{h} \int_{0}^{h} \umean_{xz}(y) \diff{y}
  \,,
\end{equation}
which lets us define a bulk flow time scale $\tscaleb=h/u_b$.

The responsiveness of particles to fluid forcing is typically measured
through the Stokes number, i.e.\ the ratio between particle and fluid
time scales. Taking the particle response time based on the Stokes drag
$\tscalep=\dpart^2\rhop/(18\nu\rhof)$, we define a first Stokes number based
on bulk time units $St_b=\tscalep/\tscaleb$ and a second Stokes number
$St^+=\tscalep/\tscalenu$ based on the
viscous time-scale $\tscalenu=\deltanu/\utau$.

Spatial fluctuations of flow quantities around the wall-parallel plane average
are defined as follows, e.g.\ for the fluid velocity:
$\mathbf{u}_f^\prime=\mathbf{u}_f-\mean{\mathbf{u}_f}_{xz}$. From these we
define second moments such as the box-averaged kinetic energy of the
fluctuations  $\ken =
\mean{\uf^\prime\cdot\uf^\prime}_{xyz}/2$,
and the plane-averaged velocity variances, e.g.\ in the streamwise
direction $\mean{u_f^\prime u_f^\prime}_{xz}$.
We will also employ time-averaging which is indicated through the
use of the operator $\mean{\,\cdot\,}_{t}$ with the averaging interval
being made more precise below.

\subsection{Numerical methods}

The numerical approach employed in this study involves three different
methodologies based on separate code frameworks. First, we use a Newton-Raphson
solver to find equilibrium solutions of the Navier-Stokes equations in plane
Couette flow. Second, the stability of the equilibrium states are tested with
the aid of a pseudo-spectral DNS solver. Third, a DNS finite-difference code
coupled with the Newton/Euler equations of rigid particle motion is employed for
time marching equilibrium solutions seeded with finite-size particles. In the
following we briefly describe the different numerical methods.

\subsubsection{Equilibrium solutions}\label{sec:computeqsol}

Equilibrium solutions are solenoidal velocity fields $\v{u}_f^{eq}$ that satisfy
the condition $\pdinline{\uf^{eq}}{t}=0$ in \cref{eq:ns_momentum} for a given
$\reynolds$, $\wllx$ and $\wllz$. The present methodology is essentially
equivalent to the one of \citet{ehrenstein:1991} which uses the Laplacian of the
wall-normal velocity perturbation and the wall-normal vorticity perturbation as
unknowns, and which employs Fourier series expansions and Chebyshev polynomials
for the spatial discretization. The resulting non-linear set of equations is
solved iteratively through a Newton-Raphson method. In order to track solutions
in the $\reynolds$, $\lambda_x$ and $\lambda_z$ parameter space, an arc-length
continuation method is employed \citep{keller:1977}. The numerical code used in
this part of the study is the same as used in \citet{Jimenez2005a}, and we refer
the reader to the latter reference for more details.

\subsubsection{Stability analysis}\label{sec:stabanalysis}

In order to investigate the stability of equilibrium solutions (in the absence
of solid particles) we perform a time-stepping-based stability analysis where
white noise is added to $\v{u}_f^{eq}$, and the disturbed velocity field is
advanced in time by a pseudo-spectral DNS code. If the disturbances are
amplified in time, the parameter point is regarded as unstable. On the other
hand, if the disturbances are damped in time, the parameter point is considered
stable. The white noise that is added to each degree of freedom is of the form
$A\exp(I\theta)$, where $I$ is the unit imaginary number, $\theta \in [0,2\pi]$
is a random phase and $A$ is the amplitude of the disturbance. The employed
numerical algorithm is as described by \citet{Kim1987}. It solves the governing
equations using again a pressure-free formulation, where the dependent variables
are the Laplacian of the wall-normal velocity and the wall-normal vorticity.
The unknowns are expanded with Fourier/Chebyshev series, and a
three-step low-storage Runge-Kutta scheme is used for time integration. The code
is essentially the same as the one used in \citet{jimenez:1999} with slight
modifications for the Couette flow boundary conditions.

\subsubsection{Interface-resolved particulate flow solver}\label{sec:fdsolver}

In order to solve the set of equations (\cref{eq:ns_mass}-\cref{eq:nl_angularmom})
for the coupled motion of fluid and particles we resort to an immersed boundary
technique. A body force term $\v{f}^{ibm}$ is introduced on the right-hand-side
of the momentum equation (\cref{eq:ns_momentum}) which is formulated such that
the no-slip condition is enforced at the particle surface. The specific
formulation is the one proposed by \citet{Uhlmann2005} which is reproduced in
appendix~\cref{apx:ibm} for convenience. A standard fractional step method is
employed, and the spatial gradients are discretized through central,
second-order finite-differences on a staggered mesh. The temporal discretization
is performed with a Crank-Nicolson scheme for the viscous terms and a three-step
Runge-Kutta scheme for advection. The particle motion
(\cref{eq:nl_linearmom}-\cref{eq:nl_angularmom}) is explicitly coupled to the
solution of the Navier-Stokes equations (\cref{eq:ns_mass}-\cref{eq:ns_momentum}).
The Poisson equation for the pseudo-pressure is solved with the aid of a
multigrid technique; the Helmholtz problems for the velocity prediction are
solved with an approximate factorization technique. The numerical code uses
three-dimensional Cartesian domain decomposition for parallelism on distributed
memory systems. The particle-wall interaction is treated with a simple repulsive
force  method \citep{glowinski:1999}, as in the DNS study of
\citet{kidanemariam:2013}; no tangential (frictional) contact force is imposed.
For the two-phase flow simulations the computational grid is uniform and
isotropic due to requirements for consistency of the immersed boundary method.
This condition is however relaxed for the single-phase simulations. The
numerical algorithm is described in detail in \citet{Uhlmann2005} and has been
used in a number of previous studies, e.g.\
\citep{uhlmann:2008,kidanemariam:2013,Uhlmann2014,chouippe:2015a}.

\section{Tracking Equilibrium Solutions}\label{sec:trackinvsol}

Before considering particulate flows, we need to establish the relevant
single-phase equilibrium solutions as well as their stability properties.

\subsection{Nagata's solutions}

We restrict our search to Nagata-like solutions as these have been extensively
studied in the literature. We consider solutions with different values of the
Reynolds number $\reynolds$, and with different fundamental wavenumbers $\wllx$
and $\wllz$. For this purpose we use two different computational domains
$\gv{\Omega}_1$ and $\gv{\Omega}_2$, the latter one being twice as large in both
the streamwise and the spanwise direction (cf.\ \cref{tb:trackeqsol}).

First we have continued one of Nagata's solutions previously used in
\citet{Jimenez2005a} with respect to the fundamental wavelengths $\wllx$ and
$\wllz$ until it matched the size of the target domains $\gv{\Omega}_1$ and
$\gv{\Omega}_2$. Then continuation in terms of $\reynolds$ was performed with
$16\times32\times16$ Fourier-Chebyshev-Fourier coefficients, and equilibrium
states of interest were recomputed with $48\times49\times48$ coefficients
($112\,896$ degrees of freedom). We have verified that this numerical resolution
is sufficient to resolve the highest coefficients of the
Fourier-Chebyshev-Fourier representation up to machine precision for all
$\reynolds$ numbers that are considered here.

\begin{figure}[p]
\captionsetup[subfigure]{labelformat=empty}
\centering
\subfloat[\label{fig:tracking_test}]{}
\includegraphics[scale=1]{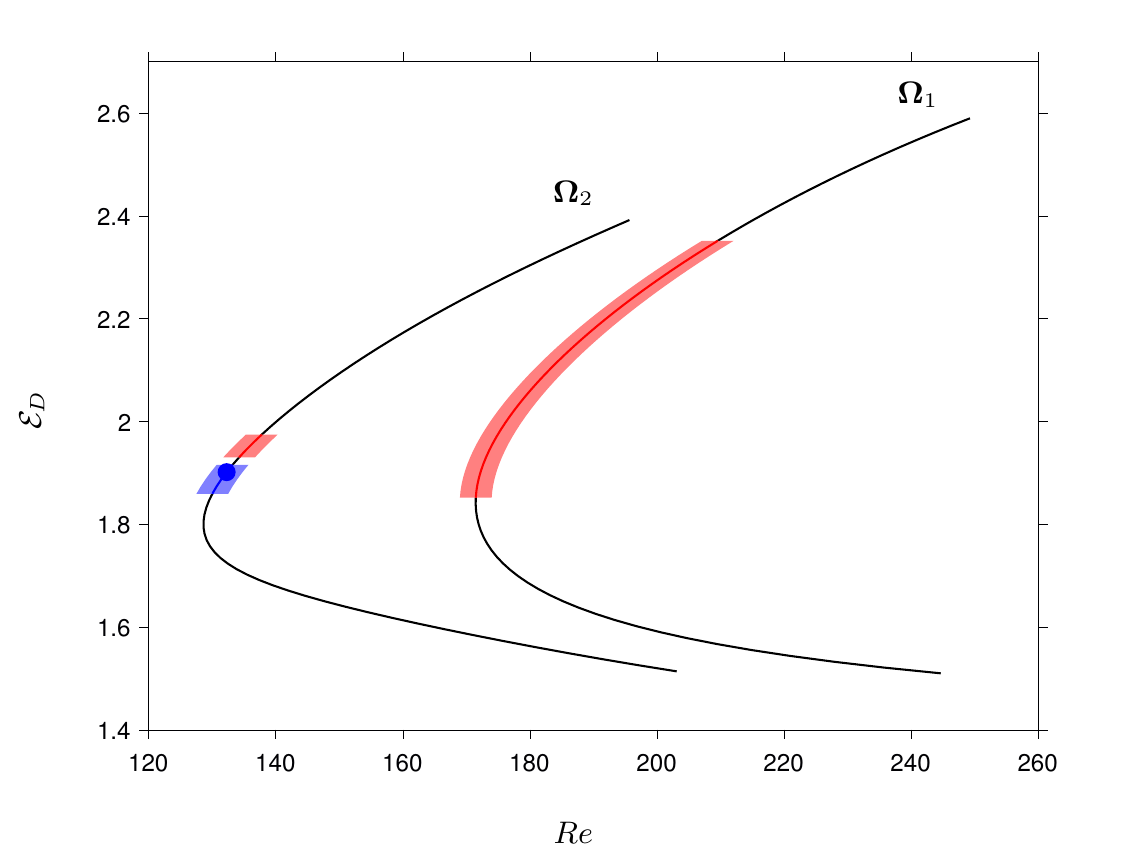}
\caption{State-space diagram for Nagata's solution in terms of the energy
dissipation rate $\dissipation$ and the Reynolds number. (a) Domain
$\gv{\Omega}_1$ with $(\wllx,\wllz)=(6h,3h)$; (b) domain $\gv{\Omega}_2$ with
$(\wllx,\wllz)=(12h,6h)$, where $\wllx$ and $\wllz$ are the fundamental
wavelengths in the streamwise and spanwise directions. The blue (red) shading
indicates stability (instability) of the solution, as determined in
\cref{{sec:stab-analysis}}. The blue dot on the upper branch of the solution in
box $\gv{\Omega}_2$ marks the selected parameter point for the present study.}
\label{fig:nagata_tracking}
\end{figure}

The average dissipation rate of the single-phase flow solutions is shown in
\cref{fig:nagata_tracking} for both sets of fundamental wavenumbers. It can be
observed that for these parameters, the branches originate at the critical
Reynolds number $171.45$ ($128.61$) in domain $\gv{\Omega}_1$ ($\gv{\Omega}_2$).
The upper branches feature substantially larger values of the dissipation rate
which indicates the presence of finer scales, i.e.\ steeper spatial velocity
gradients. Note that many previous authors have computed Nagata-type solutions
for plane Couette flow \citep[e.g.][]{waleffe:03,gibson:09}; however, we
re-computed them here for convenience and for having precise control over the
cell size. In the following we focus upon equilibrium solutions on the
upper-branch, as they are directly related to the turbulent state
\citep{Jimenez2005a}, as further discussed in \cref{sec:selecteqsol}.

\subsection{Stability analysis} \label{sec:stab-analysis}

\begin{figure}[p]
\centering
\captionsetup[subfigure]{labelformat=empty}
\centering
\subfloat[\label{fig:stabcheck_6_3_Re190p62}]{}
\subfloat[\label{fig:stabcheck_12_6_Re132p25}]{}
\includegraphics[scale=.85]{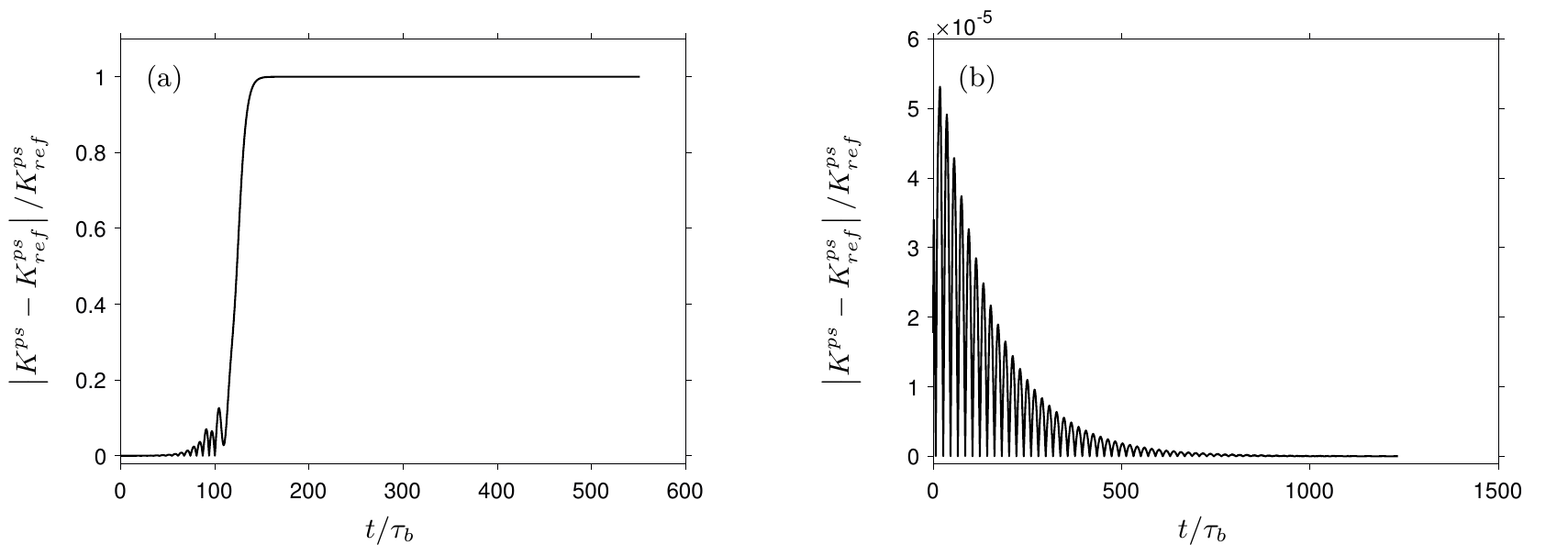}
\caption{Example of the stability analysis showing an unstable and a stable
equilibrium solution. In (a) we see that for domain $\gv{\Omega}_1$ and
$\reynolds=190.62$ the initial disturbances get amplified with time and that
after approximately $200\,\tscaleb$ the laminar base flow is recovered. In (b)
the initial disturbance in domain $\gv{\Omega}_2$ with $\reynolds=132.25$ is
continuously damped and the unperturbed case is recovered for
$t>1000\,\tscaleb$.}
\label{fig:stabcheck}
\end{figure}

In order to determine the stability of the equilibrium solutions, we added white
noise with an amplitude $A=10^{-5}\,U$ (cf.\ \cref{sec:stabanalysis}). Each of
these perturbed solutions were integrated in time using the pseudo-spectral
time-stepper, and, in total, we analyzed $10$ ($6$) parameter points on the
upper-branch of domain $\gv{\Omega}_1$ ($\gv{\Omega}_2$). The stability of each
parameter point was judged based on the box-averaged kinetic energy of the
fluctuations. Note that the superscript ``$ps$'' indicates a quantity computed
with the aid of the pseudo-spectral time-stepper, whereas the unperturbed
kinetic energy is denoted as $\kenspec_{ref}$. The cases in which the
perturbation caused the system to move away from its equilibrium value were
regarded as unstable, and otherwise as stable. As an example,
\cref{fig:stabcheck}(a) shows an unstable parameter point leading to the laminar
state, and \cref{fig:stabcheck}(b) shows a stable one. Our analysis reveals that
all the investigated parameter points of domain $\gv{\Omega}_1$ are unstable,
whereas the upper-branch of domain $\gv{\Omega}_2$ is stable within a narrow
$\reynolds$ region in the proximity of the turning point. The results of this
stability analysis are summarized by the shaded areas in
\cref{fig:nagata_tracking}, and the specific results are documented in
\cref{apx:stability}. Although, we did not perform a sensitivity study by
varying the amplitude of the initial perturbation, we have also applied the same
procedure to successfully reproduce the results of \citet{Clever1997}. In that
study, the authors show that the upper-branch of Nagata's solutions with
$(\wllx,\wllz)=(4\pi h,2\pi h)$ is indeed stable for $\reynolds$ in the vicinity
of $\reynolds_{min}$.

Note that the above analysis does not take into account possible
instability with respect to sub-harmonic perturbations, since we
perform the DNS in a domain which is commensurate with the fundamental
harmonic of Nagata's solution (i.e.\ with $L_x=\lambda_x$ and
$L_z=\lambda_z$).

\subsection{Characteristics of the selected equilibrium solution}
\label{sec:selecteqsol}

For the remainder of this work we select as our principle parameter point the
solution at $\reynolds=132.25$ on the upper branch of domain $\gv{\Omega}_2$
(cf.\ \cref{tb:nagatare132} for precise values of the physical parameters).

\begin{figure}[p]
\centering
\includegraphics[scale=1]{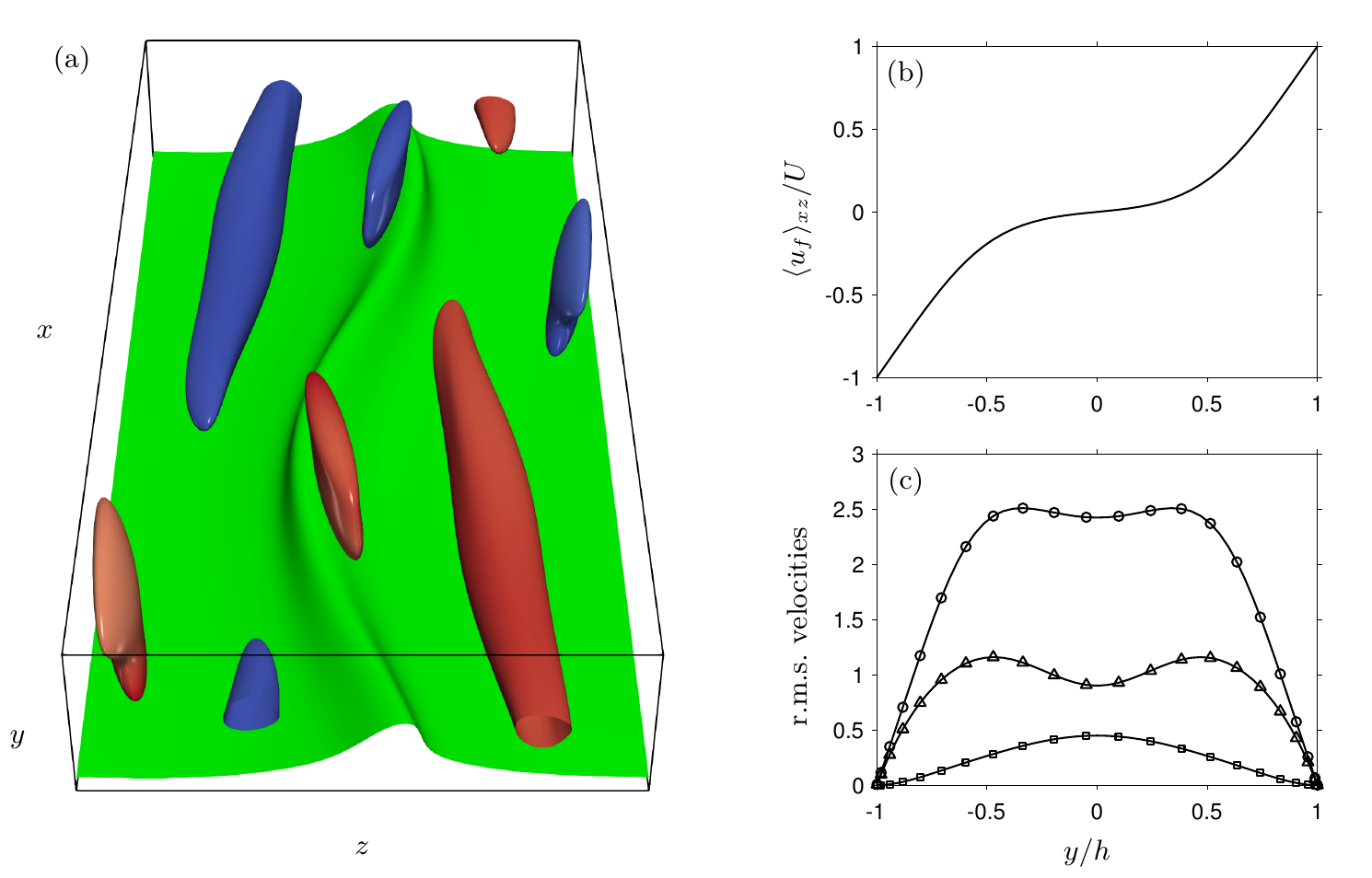}
\caption{Upper branch of Nagata's solutions at $\reynolds=132.25$ and
$(\wllx,\wllz)=(12h,6h)$. (a) Iso-surfaces: $u=\text{min}\,u(x,y=0,z)$ (green);
Q-criterion of \citet{hunt:88} with $Q=0.7\,\text{max}(Q)$, colored according to
the sign of the streamwise vorticity, i.e.\ $\omega_x<0$ (blue) and $\omega_x>0$
(red). (b) Mean streamwise velocity normalized by the wall speed. (c) Root mean
square velocities in wall units: $\urms/\utau$
\mbox{(\protect\oblacksolid)}; $\vrms/\utau$
\mbox{(\protect\sblacksolid)}; $\wrms/\utau$ \mbox{(\protect\tblacksolid)}.}
\label{fig:ubvisualization}
\end{figure}

In order to provide insight into the structure of the chosen single-phase flow
field, we show in \cref{fig:ubvisualization}(a) iso-surfaces of the Q-criterion
of \citet{hunt:88} colored according to the sign of streamwise vorticity,
plotted alongside iso-surfaces of the streamwise velocity. It can be seen that a
wavy low-speed streak is flanked by counter-rotating vortices resembling the
coherent structures typically found in near-wall turbulence. These features are
present despite the low value of the friction Reynolds number $\retau=15.87$.

The corresponding wall-normal profiles of the mean streamwise velocity
and of the standard deviations of velocity (all with respect to the
wall-parallel plane average) for this solution are shown in
\cref{fig:ubvisualization}(b,c).
It can be seen that these low-order moments of the velocity field
exhibit the qualitative features of their statistical counterparts
found in turbulent flows \citep{bech:95}, e.g.\ a representative
anisotropy and a distinct peak of the streamwise fluctuation
intensity.

\section{Exact Coherent Structures and Finite-Size Particles}
\label{sec:invsolparticles}

Let us now turn to the main part of this work by adding finite-size particles to
the previously selected single-phase solutions. Particles are inserted into the
fluid domain with their initial velocity set equal to that of the unperturbed
fluid at the initial position of the particle's centroid. In total, we have
conducted 15 independent simulations with physical parameters as listed in
\cref{tb:particleparams}. While the particle diameter is set to $\dpart/h=0.156$
($\dplus=2.48$) throughout this work, we consider a range of density ratios
$\rhop/\rhof$ in order to explore the effect of particle inertia.

The chosen number of grid nodes is $768\times128\times384$ except where stated
otherwise. The uniform isotropic grid spacing is therefore equivalent to $\Delta
x^+=\Delta y^+=\Delta z^+=0.25$ which in turn corresponds to a particle
resolution of $\dpart/\Delta x=10$. Note that we have checked that the results
do not change significantly when refining the grid by a factor of $2/3$, i.e.\
with a grid comprising $1152\times193\times576$ nodes and a particle resolution
of $\dpart/\Delta x=15$. Details on this refinement study are presented in
\cref{apx:convergence10}.

\subsection{Single particle}

Let us first consider case S10 which features a single particle with
density ratio $\rhop/\rhof=10$, corresponding to a wall-unit-based
Stokes number of $\stokesplus=3.41$.

The particle is released at an arbitrary position in the vicinity of the bottom
wall, after which the simulation is run for a total duration of
$T_{obs}=1500\,\tscaleb$. Based on the resulting data we can now address the
main questions laid out in the introduction of this study: does the presence of
a single-particle disturb the equilibrium solution significantly? Does the
particle motion exhibit preferential concentration with respect to the coherent
flow structures?

Remarkably, it turns out that the presence of the finite-size solid particle
does not disfigure the equilibrium state, and Nagata's solution is surprisingly
well preserved even after $1500\,\tscaleb$. First, in
\cref{fig:kotime_particle_s10} we compare the time evolution of kinetic energy
in case S10 with the corresponding results from the single-phase simulation
(\cref{fig:kotime_12_6_fd_vs_spec}). Here the reference single-phase solution is
represented by its time averaged value in the interval $300<t/\tscaleb<350$,
$\mean{\kenfd_{sp}}_t$ (where the subscript ``$sp$'' was added to denote
single-phase). It is observed that the signal exhibits a damped oscillation
which decays on the order of ${\cal O}(10^3)$ bulk time units, finally yielding
a sinusoidal periodic signal (cf.\ \cref{fig:kotime_particle_s10_period}). The
period of the signal in the asymptotic state is $T_{po}=2.04\,\tscaleb$, and the
amplitude of its fluctuations only measures approximately $10^{-5}$ times the
mean kinetic energy of the unperturbed flow field, $\mean{\kenfd_{sp}}_{t}$. The
discrepancy in the mean value of kinetic energy induced by the presence of the
single particle  is marginal and on the order of
$10^{-4}\mean{\kenfd_{sp}}_{t}$, thus indicating that Nagata's solution is
essentially preserved at this parameter point. This observation is also
confirmed by flow visualization, which yields largely consistent structures as
in the single-phase case (cf.\ \cref{fig:ubvisualization}), and which has
therefore been omitted here. A refined analysis shows that the introduction of
the particle triggers a very slight unsteadiness of the macroscopic flow field,
which corresponds to Nagata's solution moving in the negative $x$-direction with
a tiny propagation velocity equal to $-2.33\cdot10^{-4}u_b$. This result has
been verified under the above mentioned grid refinement.

\begin{figure}[p]
\centering
\includegraphics[scale=.9]{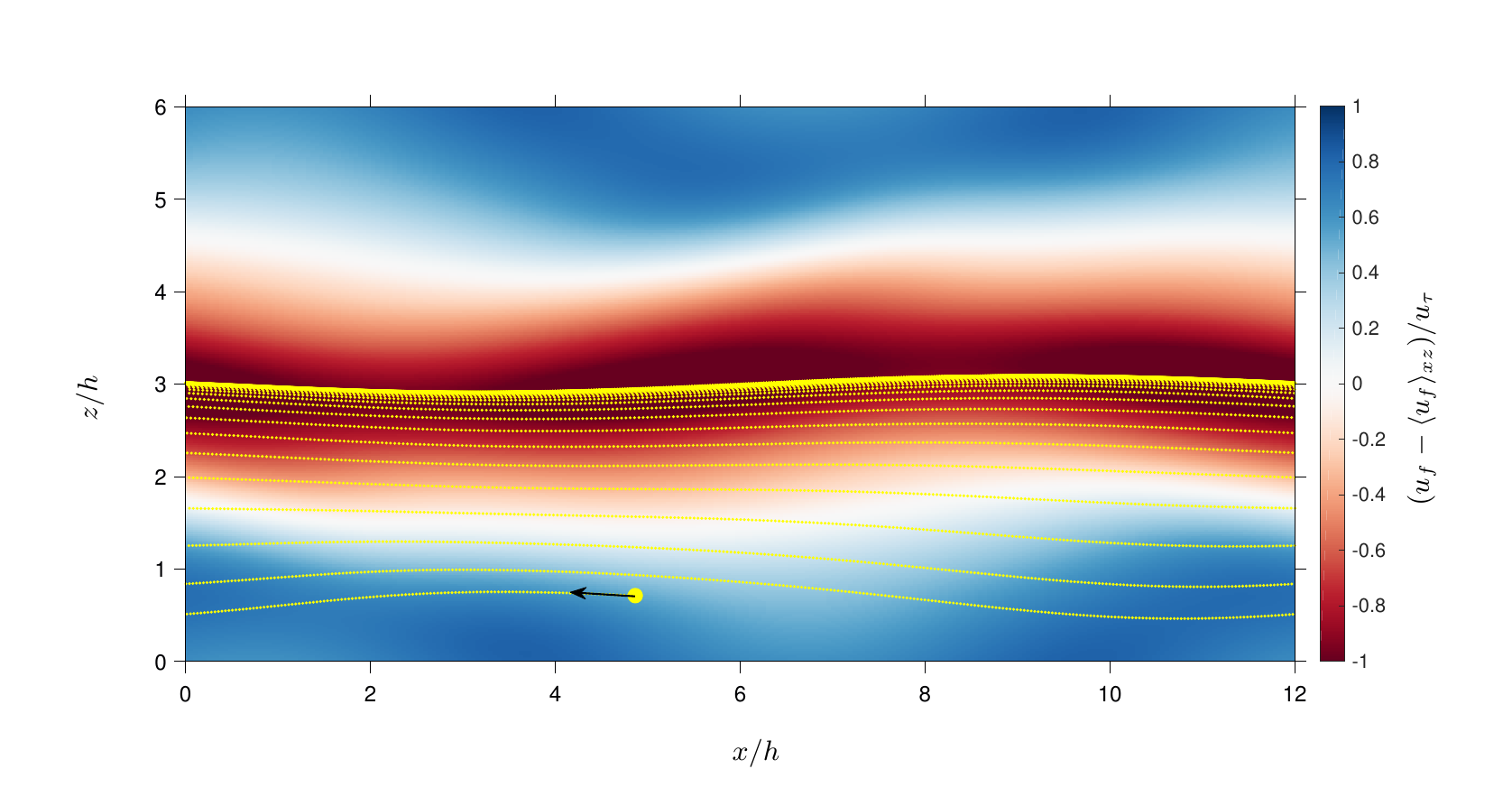}
\caption{Iso-contours of fluctuations of the streamwise fluid velocity in an
$(x,z)$-plane that intersects the particle centroid. The iso-contours evidence
the presence of a low-speed streak centered around $z/h=3$. Note that the
depicted flow field corresponds to the unperturbed flow. The yellow circle
denotes the initial particle position. The dotted line indicates the trajectory
of the particle over the course of the simulation. }
\label{fig:single_trajectory}
\end{figure}

\begin{figure}[p]
\captionsetup[subfigure]{labelformat=empty}
\centering
\includegraphics[scale=1]{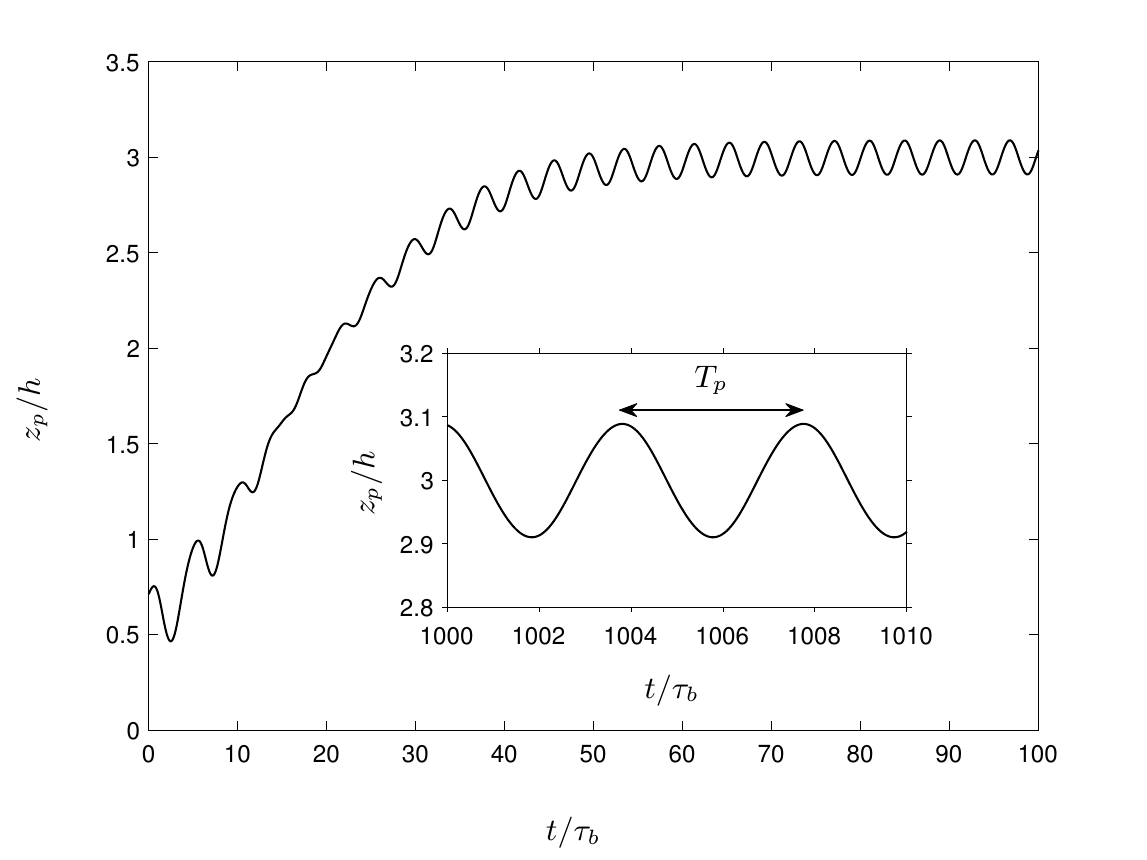}
\caption{Spanwise particle position over time in bulk units for run
S10. The inset shows a zoom of the data in the asymptotic regime, when the
motion is periodic with period $T_p=3.93\,\tscaleb$ and an amplitude of
$0.09\,h$.}
\label{fig:zpossingle}
\end{figure}

Let us now turn to the particle dynamics. We observe that the particle moves in
the (negative) streamwise direction at roughly the speed of the surrounding
fluid. Due to the action of the spanwise velocity induced by the
quasi-streamwise vortices, it slowly drifts towards the spanwise center of the
domain, approaching the low-speed fluid region (cf.\
\cref{fig:single_trajectory}). Meanwhile, the wall-normal motion of the particle
is insignificant, and it remains near its initial wall-normal position during
the course of the simulation, as expected from its large Galileo number value.
By following the particle's spanwise position (\cref{fig:zpossingle}), we
observe that it takes about $50\,\tscaleb$ for the particle to reach the region
occupied by the low-speed streak (around $\zpar/h=3$). Subsequently, the
spanwise particle position attains a state with a purely sinusoidal oscillation
with amplitude $0.09\,h$ and period $T_p=3.93\,\tscaleb$, as can be seen from
the inset in \cref{fig:zpossingle}. The average streamwise particle velocity
measures $\mean{\upar}_{ft}=-0.94\,U$, and the time for one particle
flow-through ($\domain{x}/\mean{\upar}_{ft}$) matches with the period of the
spanwise oscillatory motion $T_p$.

As a consequence of the preferential location of the particle in the low-speed
streak, it turns out that the time-average particle velocity in the asymptotic
state significantly differs from the corresponding average fluid velocity. Let
us define an apparent velocity lag as follows
\begin{equation}\label{equ-def-apparent-velocity-lag}
u_{lag}=
\mean{u_f}_{xz}(y=-h+D/2)-\mean{\upar}_{ft}
\,.
\end{equation}
Here we obtain a value of $u_{lag}=1.1\,u_\tau$ which is comparable to what has
been observed in DNS of turbulent horizontal channel flow by
\citet{kidanemariam:2013}: these authors' particles with diameter $D^+=7$ in a
flow with $Re_\tau=185$ exhibited an apparent velocity lag of approximately
$2\,\utau$ when located in the immediate vicinity of the wall. Let us underline
that the apparent lag does not reflect the relative velocity seen by the
particle, but that it is a consequence of the separate averaging of each phase
in equation (\cref{equ-def-apparent-velocity-lag}), and, therefore, that it
reflects the statistical bias due to preferential particle concentration.

It is also noteworthy that the mechanism of particle migration is caused by the
spanwise flow velocity induced by the quasi-streamwise vortices. Consequently
the spanwise particle motion is essentially due to the hydrodynamic drag force
which can be modeled by standard quasi-steady drag formulae \citep{clift:78}. It
should then in principle be possible to obtain a fair reproduction of the
present results in the context of a point-particle approach, i.e.\ without
resolving the flow around the suspended rigid particle \citep{balachandar:10},
and to use the present set-up as a testbed for such models. It appears worthwile
to further explore this avenue in future studies.

\subsection{Multiple particles}

\begin{figure}[p]
\centering
\includegraphics[scale=.9]{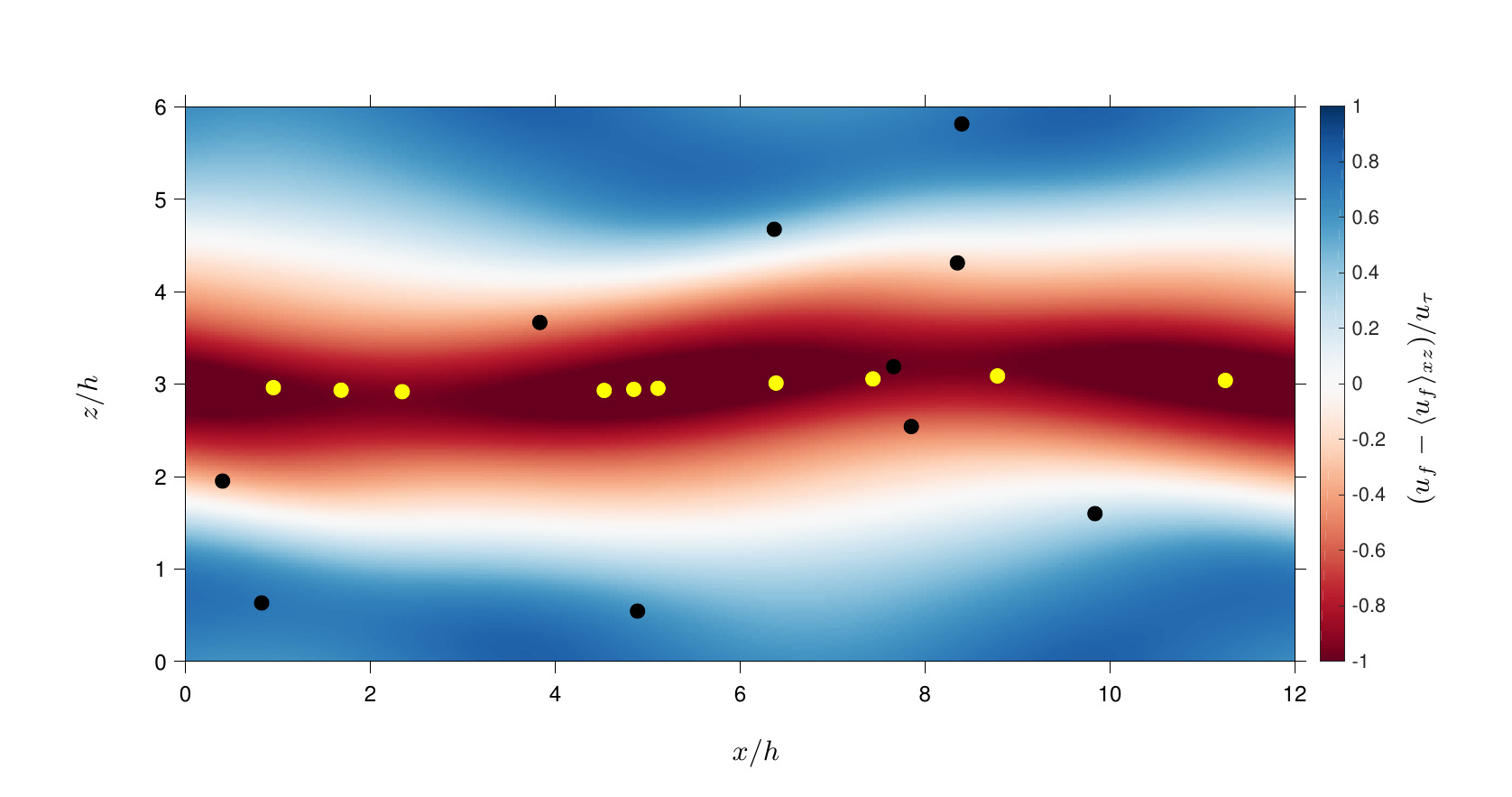}
\caption{Same graph as in \cref{fig:single_trajectory}, but showing the initial
positions (black circles) of the suspended particles in the multi-particle case
M10, as well as the particle positions after an elapsed time of $120\,\tscaleb$
(yellow circles). }
\label{fig:multpartinit}
\end{figure}

\begin{figure}[p]
\captionsetup[subfigure]{labelformat=empty}
\centering
\includegraphics[scale=1]{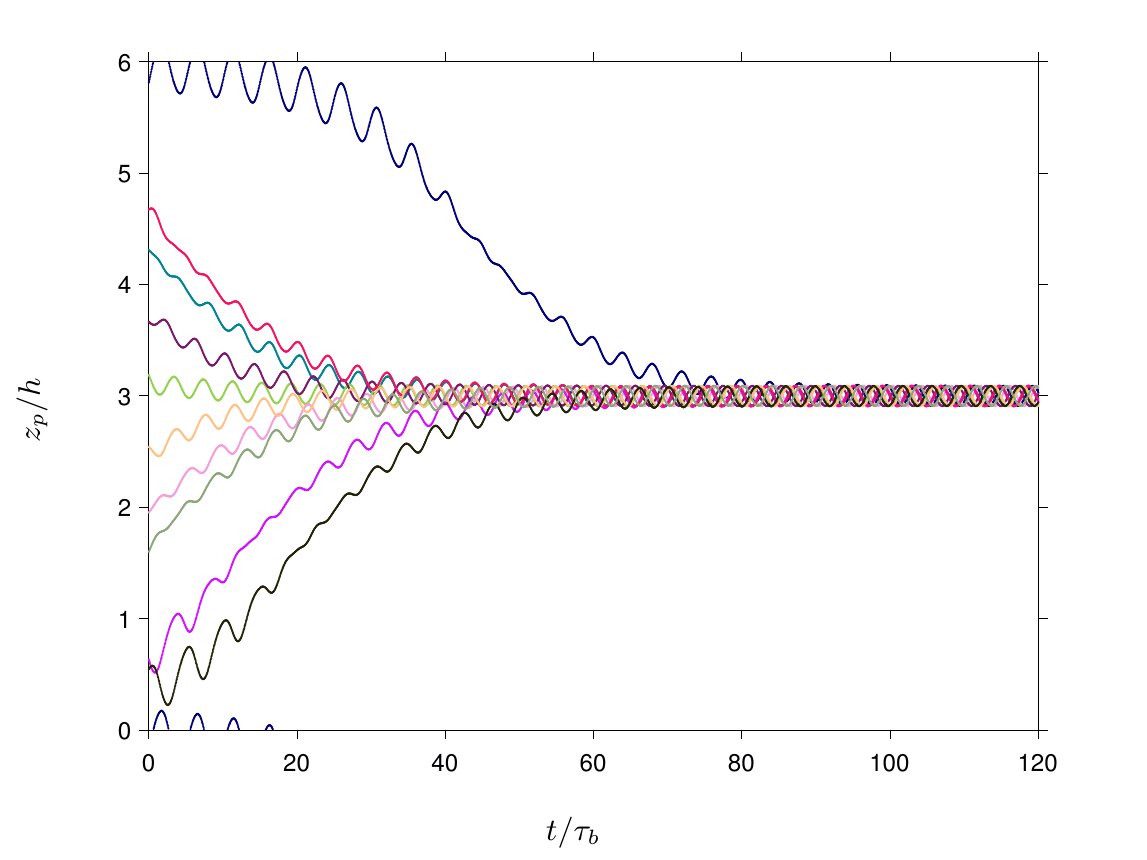}
\caption{Spanwise particle position $\zpar$ over time in bulk units
for the multi-particle case M10. Each line corresponds to a different particle.}
\label{fig:zposmulti}
\end{figure}

In order to check the dependency of our results upon the particle position at
its time of release, we have simulated the temporal evolution starting from
various initial values. For reasons of efficiency, we have done these tests in a
multi-particle simulation (case M10 in \cref{tb:particleparams}), where 10
particles are simultaneously released at random positions in the wall-parallel
plane in the vicinity of the lower  wall (cf.\ \cref{fig:multpartinit}). The
evolution of the spanwise position, which is shown in \cref{fig:zposmulti},
reveals that the spanwise motion in the asymptotic state is harmonic with the
same period and amplitude for all particles. The latter matches well with the
values found in the single-particle case S10, demonstrating two points: first,
that particles migrate towards the low-speed streak irrespective of their
initial position, and, second, that collective effects are not felt at this low
particle concentration. Therefore, this multi-particle simulation confirms the
observation that the persistent low-speed streak acts as a stable attractor to
near-wall particle motion.

\subsection{Constrained particles}

While dynamical simulations (as the ones which we have presented above) can be
used for the detection of stable equilibria of the system, it is not easy to
find possible unstable equilibria with this method. Furthermore, it is of
general interest to determine the stability properties of particle locations
with respect to the coherent structures of the background flow in more detail.
In order to obtain such additional information, it is useful to resort to the
method of constrained simulations, where some otherwise dynamical quantity is
held fixed, while the applied constraining force (or torque) is measured. This
approach has been successfully used e.g.\ by \citet{patankar:01b} and by
\citet{joseph:02} for the investigation of the wall-normal migration of
neutrally-buoyant particles in wall-bounded shear flows.

In the present context we proceed as follows. We seed the equilibrium solution
with particles whose motion in the spanwise direction is suppressed, while they
are free to move in the streamwise/wall-normal plane. We then let the simulation
evolve until an asymptotic state is reached, which now features the particle
translating on an essentially straight path (in the $x$-direction) in a
time-periodic regime of motion, during which the hydrodynamic force acting upon
the particles oscillates with the same period. The force component in the
constraining direction at a given spanwise position, $F_z(z,t)$, is our primary
quantity of interest. In case of a steady state system, the equilibria are then
readily obtained as the positions $z_e$ where the force $F_z(z_e)$ vanishes;
stability (instability) can be detected by checking for negative (positive)
gradients $\partial F_z/\partial z$ at the equilibrium positions.
In the present case, where the constraining force is time-dependent, we perform
the same analysis for the time-average of the spanwise component of the particle
force, $\langle F_z(z,t)\rangle_t$. It should be noted that this is clearly a
simplification, as the effect of temporal fluctuations upon the stability of the
spanwise particle position are thereby effectively ignored. Therefore, this
point might merit further investigation in future studies.

\begin{figure}[p]
\centering
\includegraphics[scale=1]{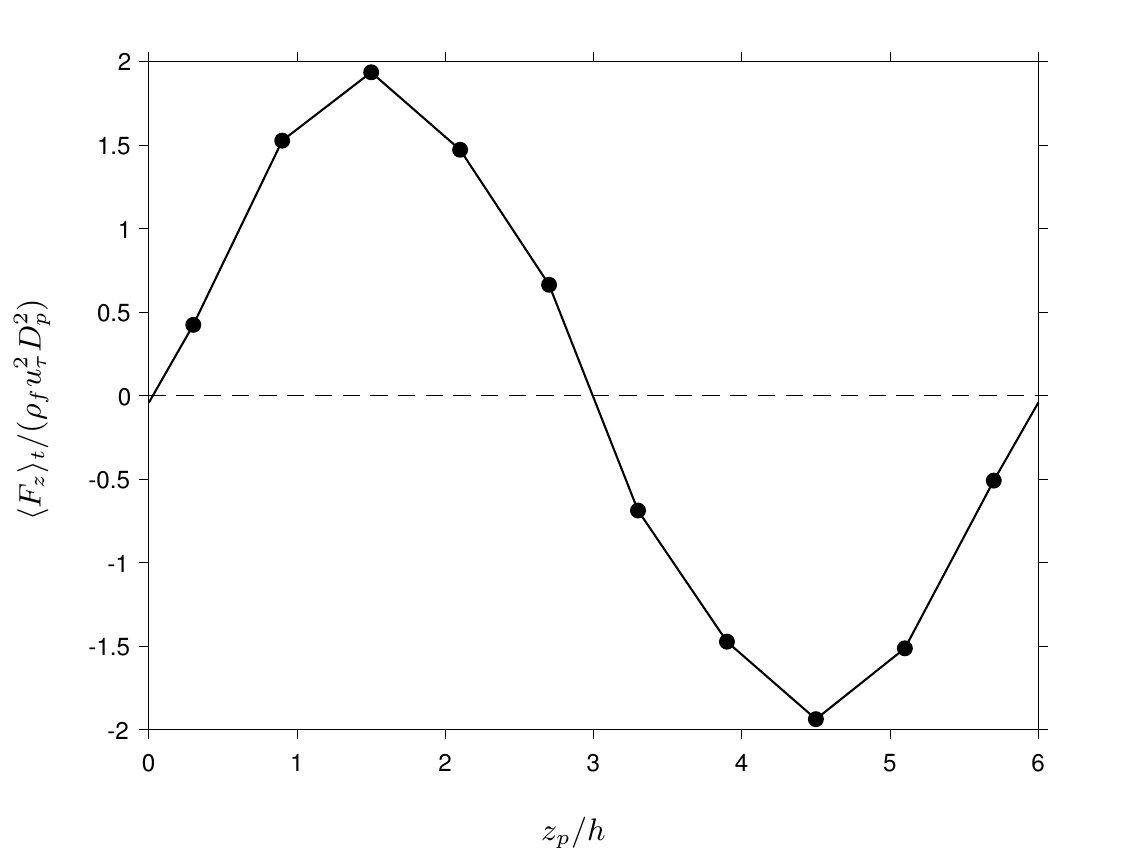}
\caption{Time-average spanwise force for the last particle cycle in
the constrained particle case C10, plotted as a function of the imposed spanwise
position (i.e.\ one datum per particle).
}
\label{fig:forceavg_constrained}
\end{figure}

In case C10 (with otherwise same parameters as cases S10 and M10, cf.\
\cref{tb:particleparams}) we perform a constrained simulation with 10 particles
simultaneously (again for reasons of efficiency). In order to minimize their
mutual influence, we initially place the particles uniformly on a diagonal in
the $(x,z)$-plane. This numerical set-up was integrated in time for
$116\,\tscaleb$, which corresponds to more than 20 passages through the domain.
However, we have verified that the forces acting on the particles already attain
their asymptotic time evolution after only two of these passages, after which
the values fluctuate by less than 5\% from passage to passage. The time-average
spanwise force $\langle F_z\rangle_t$ acting on the particles (in the asymptotic
regime) is shown in \cref{fig:forceavg_constrained} as a function of the
respective spanwise position. As expected, the force exhibits two zero-crossings
at $z_e/h=\{3,\,6\}$. One equilibrium position coincides with a positive
gradient of the spanwise force ($z_e/h=6$), and it can therefore be classified
as unstable; the other one ($z_e/h=3$), which coincides with the mean location
of the low-speed streak, is stable, since the gradient of $\langle F_z\rangle_t$
is negative at that position.

Therefore, this simplified analysis confirms that the low-speed region is the
only stable equilibrium for the particles' spanwise location. Please note that
this constrained simulation is significantly more efficient than the
unconstrained one, since the transient time interval to be covered is much
shorter. The present results consequently underline that the constrained
simulation approach is a very useful tool in the context of computationally
demanding resolved-particle DNS.

\subsection{Density ratio}

In order to elucidate the scaling of the particle mobility as a
function of its inertia, we have varied the solid-to-fluid density ratio
by a factor of twenty from case S04 to S80 (cf.\
\cref{tb:particleparams}), while maintaining all remaining physical and
numerical parameters fixed (the Stokes and Galileo numbers
vary accordingly).
Please note that here we strictly separate the inertia effect from
the geometrical size effect by keeping the particle diameter fixed.
This is fundamentally different from varying the Stokes number in
the context of a point-particle approach.

\begin{figure}[p]
\centering
\captionsetup[subfigure]{labelformat=empty}
\subfloat[\label{fig:spanwiseposdratio}]{}
\subfloat[\label{fig:spanwise_velocity_dratio}]{}
\includegraphics[scale=.85]{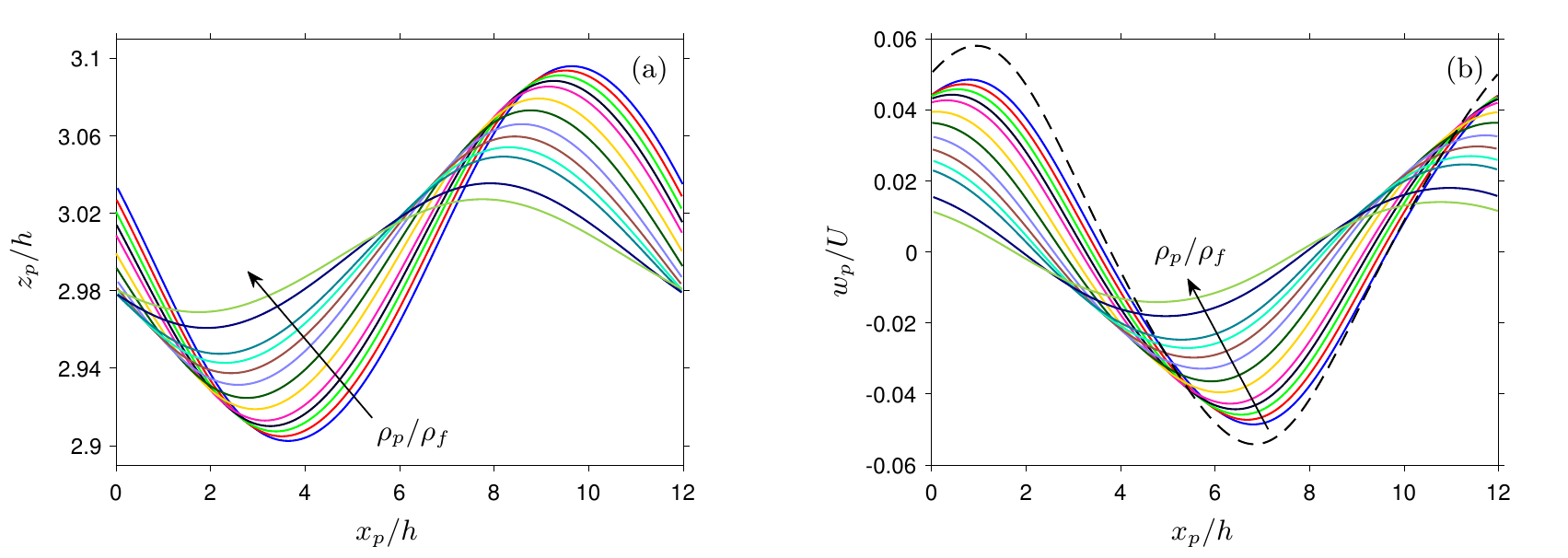}
\caption{Single particle simulations at various solid/fluid density ratios. The
data is for the asymptotic regime with fully-developed periodic particle motion.
(a) Spanwise particle position as a function of the streamwise position. (b)
Spanwise particle velocity as a function of the streamwise position. The
color-code in both graphs refers to each run S04-S80 in
\cref{tb:particleparams}, where the arrow indicates the direction of increase of
$\rhop/\rhof$. The dashed curve in (b) shows the spanwise fluid velocity of the
unperturbed flow field at the mean low-speed streak location and at a
wall-distance equivalent to the particle's centroid.}
\label{fig:spanwise_pos_velo_dratio}
\end{figure}

\begin{figure}[p]
\centering
\captionsetup[subfigure]{labelformat=empty}
\subfloat[\label{fig:amplitude_zpos_dratio}]{}
\subfloat[\label{fig:amplitude_wpar_dratio}]{}
\includegraphics[scale=.85]{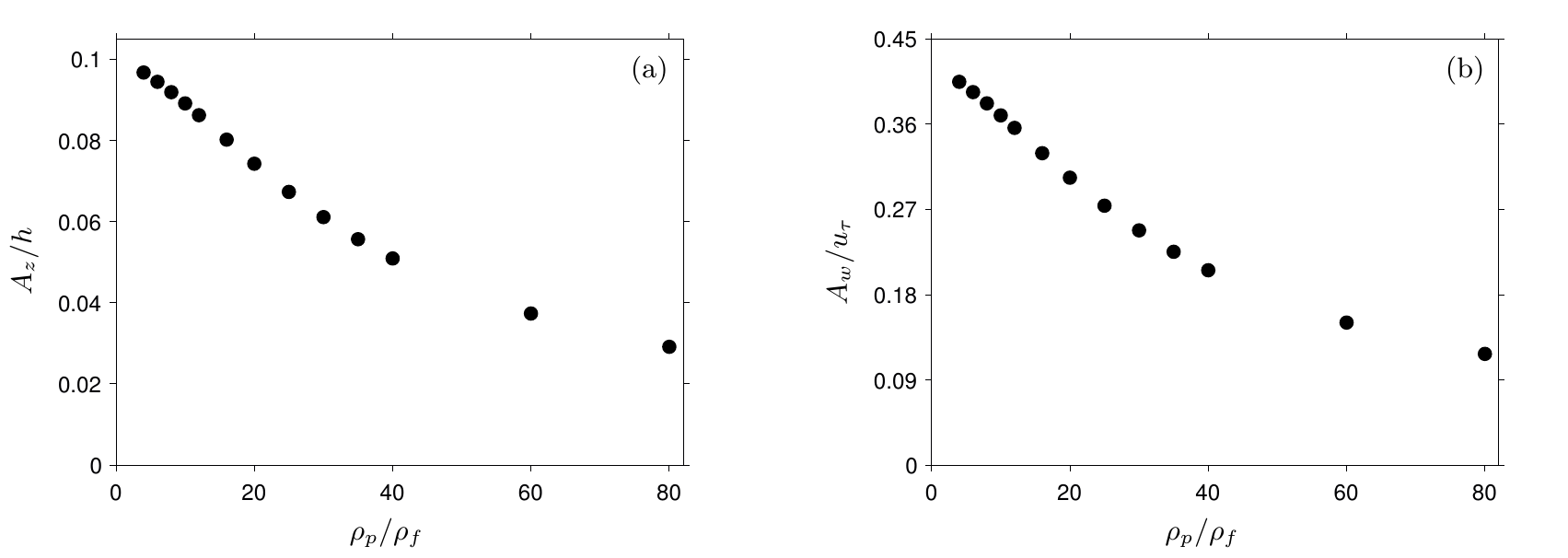}
\caption{Amplitude of the spanwise particle motion (a) and the
spanwise particle velocity (b) as a function of density ratio for cases S04-S80.
}
\label{fig:amplitude_dratio}
\end{figure}

In all runs, the initial position of the particle is the same as in case S10,
and the simulations are carried out for $110\,\tscaleb$. After their initial
release, all cases display similar dynamics as case S10 (as already shown in
\cref{fig:zpossingle}), and the time for the particles to reach the low-speed
streak does not appear to depend significantly on the density ratio
$\rhop/\rhof$ (figure omitted). Note that in all cases the particle remains in
contact with the lower wall at all times, except for a short initial transient
at the lowest Galileo number case S04. At later times, all runs exhibit a
sinusoidal periodic motion in the $(x,z)$-plane, analogous to case S10, as can
be seen from \cref{fig:spanwise_pos_velo_dratio}. In the asymptotic state, the
particles oscillate around the same time-average position
$\langle\zpar\rangle_t/h=3$, irrespective of their density ratio. The amplitude
and the phase of the particle motion, however, vary monotonously with
$\rhop/\rhof$. More specifically, these amplitudes ($A_z$ for the spanwise
particle excursions, and $A_w$ for the spanwise particle velocity fluctuations)
are defined as one half of the difference between maximum and minimum values of
the data shown in \cref{fig:spanwiseposdratio,fig:spanwise_velocity_dratio},
respectively. \cref{fig:amplitude_dratio} then shows that over the investigated
parameter range both amplitudes (spanwise particle position and velocity) first
decrease roughly linearly with the density ratio, and then exhibit an
increasingly non-linear evolution with $\rho_p/\rho_f$. This behavior can be
qualitatively explained with the aid of very simple point-particle arguments,
invoking Stokes drag to be the only particle force acting in the spanwise
direction and assuming that the velocity seen by the particle is the unperturbed
flow velocity of the equilibrium solution. Under these assumptions the amplitude
of the spanwise particle excursions and of its spanwise velocity fluctuations
decays as $(\rho_p/\rho_f)^{-1}$ for large density ratios. However, in order to
allow for a quantitative match with the data for finite-size particles, a more
elaborate model would be required, taking into account unsteady force terms,
finite-size and finite-Reynolds number corrections as well as wall effects.
Again, the formulation of a realistic force model in the context of a
point-particle approach would be a rewarding subject for future research, for
which the present configuration can serve as a useful validation case.

\section{Conclusions and perspectives for further studies}

In this work we have attacked the problem of fluid-particle interaction with the
aid of particle-resolved DNS based on an immersed-boundary method. Instead of
considering turbulent flow and performing statistical analysis we have used a
non-trivial equilibrium solution which features exact coherent structures
(streaks and vortices) representative of the principal ingredients of
wall-bounded shear flows. The present strategy has the advantage of a much
reduced complexity, in particular in terms of ease of data-analysis.

More specifically, we have focused here on the upper branch of Nagata's solution
for plane Couette flow, which has previously been shown to reproduce some of the
low-order statistics of the turbulent flow state. For the present work we have
restricted our attention to a low Reynolds number, for which the single-phase
flow solution is stable to finite-amplitude white noise perturbations. We have
found that adding either a single or a small number of heavy spherical particles
with a diameter equivalent to one twelfth of the gap width (2.5 wall units) does
not significantly alter the flow structure, such that the background flow is
essentially maintained. At the same time, a solid particle, which due to gravity
remains in a plane adjacent to the lower horizontal wall, migrates towards the
region occupied by a low-speed streak, and then attains a regime of periodic
motion which is independent of the initial position. As a consequence of the
particle's preferential location, it does not sample the flow field uniformly,
and, therefore, its time-average velocity differs from the (unperturbed) mean
flow velocity at the wall-distance of its centroid.

This apparent velocity lag has previously been reported in experiments and in
DNS studies of horizontal channel flow in the turbulent regime. Although past
studies have already laid out the above mechanism leading to preferential
particle concentration and its consequences, the data-analysis (involving
coherent structure eduction and particle-conditioned statistical averaging) has
required much effort. When using exact coherent structures as the background
flow, this effort is significantly reduced. Furthermore, the technique of
simulating constrained particles in order to determine the stability properties
of particle motion can be applied to flow fields which are invariant solutions
to the Navier-Stokes equations, as has been demonstrated in the present work.

Several new physical results have been obtained here by performing a sweep of
the solid-to-fluid density ratio ($\rhop/\rhof=4\ldots80$ corresponding to
$St^+=1.4\ldots27$). It turns out that the time it takes for the particle to
migrate to the equilibrium position (inside the low-speed streak) does not
significantly depend on the particle density in the range under investigation.
In the asymptotic regime, on the other hand, the effect of particle inertia
leads to the amplitude of the oscillations of the spanwise particle motion to
decrease non-linearly with particle density.

To sum up, the present work demonstrates that it is technically feasible and
beneficial to study invariant solutions (and in particular equilibrium
solutions) with suspended finite-size particles. This set-up provides a
numerical laboratory with minimal complexity, yet still relevant to the full
problem of sustained turbulence. We expect the proposed approach to be fruitful
in future studies on various aspects of particulate flow, and some perspectives
for future work are briefly discussed in the following.

The most immediate avenue to pursue is to expand the study in parameter space.
First, we note that in the present contribution we have not varied the particle
diameter. It is obviously of interest to determine the limit for the occurrence
of the preferential location mechanism as the particle size increases with
respect to the scale of the coherent structures. Varying particle diameter and
density independently will help to distinguish between relative size effects and
inertial effects, as an alternative to the amalgamation of both in form of the
Stokes number. Other important parameters which are straightforward to
investigate in this framework include the role of gravity (sweep of Galileo
number, orientation of the vector of gravitational acceleration), as well as an
investigation of collective effects.

One important aspect which merits further analysis is the consideration of
parameter points at which the equilibrium solution as such is already unstable
(e.g.\ Nagata's upper branch solution at larger Reynolds number). Preliminary
computations indicate that the present approach still makes sense, as long as
the solution remains suitably close to the equilibrium point over an interval
which is long compared to the particle's characteristic time scale. For some
parameter points it is expected that the presence of particles adds to the
de-stabilization of the flow. The study of the precise mechanisms by which
particles contribute to the enhancement (or attenuation) of instability can be a
rewarding subject by itself. Along these lines, we expect that the study of
invariant solutions and finite-size particles can make a valuable contribution
to the elucidation of open questions in the area of transition to turbulence in
particulate flows \citep{matas:03,loisel:13}.

Since steady equilibrium solutions, such as the one considered herein, as well
as other traveling-wave type solutions do not involve a notion of characteristic
life-time of the coherent structures (i.e.\ the streaks and vortices are always
present), possible effects of disparity of proper time scales in the
fluid/particle interaction problem are not represented in this system.
Therefore, it would be of great value to consider time-periodic invariant
solutions (i.e.\ periodic orbits) in future studies involving solid particles.
Prime candidates in this direction are the periodic solutions for plane Couette
flow discovered by \citet{Kawahara2001a}, which feature a self-sustaining cycle
involving velocity-streak break-up and regeneration of quasi-streamwise
vortices. At the present time, however, it is not clear whether it will be
feasible to conduct numerical experiments involving finite-size particles
suspended in such periodic orbit solutions.

Finally, let us mention that the present approach can obviously be extended to
the study of small particles whose near-field does not need to be resolved
(i.e.\ idealized as point particles). If the disperse phase is sufficiently
dilute, the point-particle study can be conducted in one-way coupled fashion
(i.e.\ the perturbation of the fluid phase can be neglected), which then renders
the numerical simulations extremely efficient. Preliminary simulations of this
type indicate that they will yield interesting information on preferential
concentration and clustering. In addition, we see an opportunity to validate and
eventually improve force models in the context of the point-particle approach
with the aid of the present set-up. Such a study would have the clear advantage
of full reproducibility. Further elaboration of this topic will be left for
future work.

\section*{Acknowledgments}

The simulations were partially performed at SCC Karlsruhe. The computer
resources, technical expertise and assistance provided by this center are
thankfully acknowledged.
\bibliographystyle{jfm_mod}
\bibliography{manuscript}

\clearpage
\appendix

\section{Algorithm of the immersed boundary method}
\label{apx:ibm}
For each time step which advances the solution from $t^n$ to $t^{n+1}$ we
perform three Runge-Kutta sub-steps with indices $\kappa=1,2,3$. The fractional
step method first computes a pre-predicted velocity field $\tilde{\mathbf{u}}$
which does not include any explicit effect of the suspended particles. Next a
force field $\mathbf{f}^{(ibm)}$ is computed which serves to impose the desired
no-slip velocity at the particle surfaces. The momentum equations are then
solved with the added force field to yield a predicted velocity field
$\mathbf{u}^\ast$. Subsequently a Poisson equation is solved for the
pseudo-pressure $\phi$ which then serves to project the velocity field upon the
divergence-free space, yielding the final field $\mathbf{u}$. The particle
equations of motion are finally updated to yield the new positions and particle
velocities.

The overall algorithm for a single Runge-Kutta sub-step with index
$\kappa$ can be written as follows:
\begin{subequations}\label{equ-hybrid-discr-algo2}
\begin{eqnarray}\label{equ-hybrid-discr-algo2-lag-rhs}
\tilde{\mathbf{u}}&=&
\mathbf{u}^{\kappa-1}
 +\Delta t
    \vphantom{\left.\mathbf{u}\right)^{\kappa-1}}
2\alpha_\kappa\nu\nabla^2\mathbf{u}^{\kappa-1}
-2\alpha_\kappa\nabla p^{\kappa-1}
   -\gamma_\kappa\left((\mathbf{u}\cdot\nabla)\mathbf{u}
   \right)^{\kappa-1} -\zeta_\kappa\left((\mathbf{u}\cdot\nabla)\mathbf{u}
   \right)^{\kappa-2}
\,,
  \\
  \label{equ-hybrid-discr-algo2-lag-interpol}
\tilde{{U}}_\beta(\mathbf{X}_l^{(m)})&=&
\sum_{ijk}\tilde{{u}}_\beta(\mathbf{x}_{ijk}^{(\beta)})\,
\delta_h(\mathbf{x}_{ijk}^{(\beta)}-\mathbf{X}_l^{(m)}(t^{\kappa-1}))\,\Delta
x^3 \,,
\quad\forall\,l;\,m;\,\beta
\\
\label{equ-hybrid-discr-algo2-lag-force}
\mathbf{F}(\mathbf{X}_l^{(m)})&=&
\frac{\mathbf{U}^{(d)}(\mathbf{X}_l^{(m)},t^{\kappa-1})
-\tilde{\mathbf{U}}(\mathbf{X}_l^{(m)},t^{\kappa-1})}{\Delta t}
\,,
\quad\qquad\qquad\forall\,l;\,m
\\
\label{equ-hybrid-discr-algo2-eul-force}
{f}_\beta^{(ibm),\kappa}(\mathbf{x}_{ijk}^{(\beta)})&=&
\sum_{m=1}^{N_p}
\sum_{l=1}^{N_L}{F}_\beta(\mathbf{X}_l^{(m)})\,
\delta_h(\mathbf{x}_{ijk}^{(\beta)}-\mathbf{X}_l^{(m)}(t^{\kappa-1}))\,\Delta
V_l^{(m)} \,,
\quad\forall\,\beta;\,i;\,j;\,k
\\
\label{equ-hybrid-discr-algo2-predict}
\nabla^2\mathbf{u}^\ast-\frac{\mathbf{u}^\ast}{\alpha_\kappa\nu\Delta t}&=&
-\frac{1}{\nu\alpha_\kappa}\left(\frac{\tilde{\mathbf{u}}}{\Delta t}
+\mathbf{f}^{(ibm),\kappa}
\right)
+\nabla^2\mathbf{u}^{\kappa-1}
\,,
\\\label{equ-hybrid-discr-algo2-poisson}
\nabla^2\phi&=&\frac{\nabla\cdot\mathbf{u}^\ast}{2\alpha_\kappa\Delta t}\,,
\\\label{equ-hybrid-discr-algo2-update-vel}
\mathbf{u}^{\kappa}&=&\mathbf{u}^\ast-2\alpha_\kappa\Delta
t\nabla\phi\,,
\\\label{equ-hybrid-discr-algo2-update-press}
p^{\kappa}&=&p^{\kappa-1}+\phi-\alpha_\kappa\Delta t\,\nu\nabla^2\phi
 \,,
  \\\label{equ-particles-newton-2-present-translation-discrete-u-DEM}
  \frac{\mathbf{u}_p^{\kappa,\,(m)}-\mathbf{u}_p^{\kappa-1,\,(m)}}{\Delta t}
                  &=&
                      \frac{\rho_f}{V_p(\rho_p-\rho_f)}\left(
                      -{\mathcal F}^{\kappa,\,(m)}
                      +\sum_{l\neq m}\mathbf{F}_{rep}^{\kappa-1,\,(l,m)}
                      +\mathbf{F}_{wall}^{\kappa-1,\,(m)}
                      \right)
                      +2\alpha_\kappa\mathbf{g}
       \,,
        \qquad\forall\,m
    \\\label{equ-particles-newton-2-present-translation-discrete-x-DEM}
    \frac{\mathbf{x}_p^{\kappa,\,(m)}-\mathbf{x}_p^{\kappa-1,\,(m)}}{\Delta t}
    &=&
    \alpha_\kappa\left(
      \mathbf{u}_p^{\kappa,\,(m)}+\mathbf{u}_p^{\kappa-1,\,(m)}
        \right)
        \,,
        \qquad\forall\,m
    \\\label{equ-particles-newton-2-present-translation-discrete-om-approx-DEM}
\frac{\boldsymbol{\omega}_p^{\kappa,\,(m)}-\boldsymbol{\omega}_p^{\kappa-1,\,(m)}}{\Delta
    t}
    &=&
        -\frac{\rho_f}{\rho_p-\rho_f}\frac{1}{(I_p/\rho_p)}\,{\mathcal T}^{\kappa,\,(m)}
        \,,
        \qquad\forall\,m
\end{eqnarray}
\end{subequations}
where $m$ is the index of a given particle ($1\leq m\leq N_p$), $\beta$ denotes
a spatial direction ($1\leq\beta\leq3$), $\mathbf{X}_l^{(m)}$ is the position of
a Lagrangian force point with index $l$ (where $1\leq l\leq N_l$) attached to
the $m$th particle, $\delta_h$ is the discrete delta function of
\citet{roma:99}, $\mathbf{x}_{ijk}^{(\beta)}$ is the position vector of a node
of the staggered Cartesian fluid grid of the velocity component in the $x_\beta$
direction with index triplet ``$ijk$'', $\tilde{{U}}_\beta$ is the velocity in
the $x_\beta$-direction interpolated to a Lagrangian position,
$\mathbf{U}^{(d)}(\mathbf{X}_l)$ is the solid body velocity of the Lagrangian
force point, $\mathbf{F}(\mathbf{X}_l^{(m)})$ is the immersed boundary force at
a Lagrangian force point, $\Delta V_l^{(m)}$ is the forcing volume associated to
the $l$th Lagrangian forcing point of the $m$th particle (equal to $\Delta x^3$
here), $\mathbf{x}_p^{\kappa,\,(m)}$ is the centroid position of the $m$th
particle, ${\mathcal F}^{\kappa,\,(m)}$ is the hydrodynamic force computed from
the sum of the immersed boundary contributions of the $m$th particle, ${\mathcal
T}^{\kappa,\,(m)}$ is the analogous hydrodynamic torque contribution,
$\mathbf{F}_{rep}^{\kappa-1,\,(l,m)}$ and $\mathbf{F}_{wall}^{\kappa-1,\,(m)}$
are the force contributions from particle-particle and particle-wall contact,
respectively. The set of coefficients $\alpha_k$, $\gamma_k$, $\xi_k$ for a
low-storage scheme leading to second-order temporal accuracy has been given in
\citet{rai:91}. Please refer to the original publication \citet{Uhlmann2005} for
more details on the algorithm.

\section{Stability analysis}
\label{apx:stability}

The stable/unstable parameter points of the upper-branch of Nagata's solutions
for domains $\gv{\Omega}_1$ and $\gv{\Omega_2}$ were previously summarized in
form of shaded areas in the  $(\reynolds,\dissipation)$ state-space diagram
(\cref{fig:nagata_tracking}). Here, we list the discrete $\reynolds$ and
$\dissipation$ values for which the stability analysis was
performed\,\textemdash\, see \cref{tb:stable_unstable}.

In \cref{tb:stable_unstable}, the reader also finds the stable/unstable
parameters points for a third domain, namely $\gv{\Omega}_3=[0,4\pi
h]\times[-h,h]\times[0,2\pi h]$. Domain $\gv{\Omega}_3$ is identical to the one
used in \citet{Clever1997}, where the authors conducted a linear stability
analysis. The reproduction of their results served to validate our stability
analysis approach, which is instead based on a time-stepper applied to an
initial field which is perturbed by white noise, as detailed in
\cref{sec:stabanalysis}. In all cases, we restrict our analysis to parameter
points located on the upper-branch of Nagata's solutions, as these are
physically more interesting for the purpose of the current work.

\section{Computing single-phase equilibrium solutions with a finite-difference
method} \label{sec:fd-single-phase-sol}

\begin{figure}[p]
\centering
\includegraphics[scale=1]{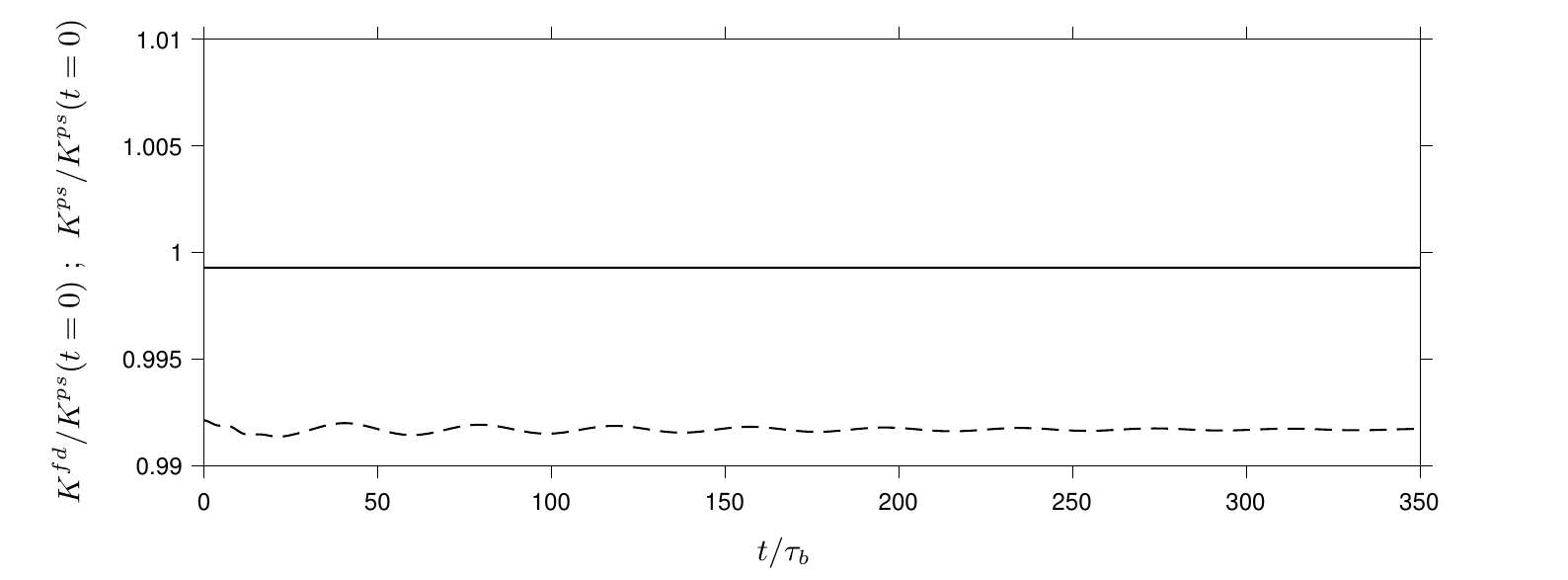}
\caption{Time evolution of the box-averaged kinetic energy of the fluctuations
for Nagata's solution with $(\wllx,\wllz)=(12h,6h)$ and $\reynolds=132.25$ when
integrated using a {pseudo-spectral time-stepper} as described in
\cref{sec:stabanalysis} (\mbox{\protect\blackline}) and with the second-order
finite-difference solver as described in \cref{sec:fdsolver} with a uniform grid
featuring widths of $(\Delta x^+,\Delta y^+,\Delta z^+)=(0.50,0.25,0.50)$,
starting from a spectrally interpolated solution at time $t=0$
(\mbox{\protect\blackdashed}).}
\label{fig:kotime_12_6_fd_vs_spec}
\end{figure}

Here we wish to verify how closely the finite-difference solution on a given
grid resembles the reference spectral results in the absence of particles. The
equilibrium solution obtained with the Newton-Raphson approach and
spectral-discretization is first spectrally interpolated upon the
finite-difference grid, before starting time-stepping.
\cref{fig:kotime_12_6_fd_vs_spec} depicts the temporal evolution of kinetic
energy for the stable parameter point (cf.\ \cref{tb:nagatare132}) when using
$384\times128\times192$ grid points, which implies a grid resolution of $(\Delta
x^+,\Delta y^+,\Delta z^+)=(0.50,0.25,0.50)$. The figure shows that the initial
difference between $\kenfd$ and $\kenspec$ (i.e.\ the error due to spatial
interpolation with a second-order method) is less than $0.01\,\kenspec$. In
\cref{fig:kotime_12_6_fd_vs_spec} a mild temporal evolution in the form of a
low-amplitude damped oscillation is subsequently observed for the
finite-difference solution, which converges to within $0.7\%$ discrepancy of the
spectral reference data.

\section{Grid convergence study for case S10}
\label{apx:convergence10}

Here we compare the main findings of case S10 with results from case S10-F which
features a refined grid. For this additional run, the number of grid points was
increased by a factor of $3/2$ yielding $1152\times$$ 193 \times 576$ nodes,
i.e.\ $\Delta x^+=\Delta y^+=\Delta z^+=0.167$, and a particle resolution of
$\dpart/\Delta x=15$. The remaining numerical and physical parameters were held
constant and are readily found in \cref{tb:particleparams}.

\begin{figure}[p]
\centering
\captionsetup[subfigure]{labelformat=empty}
\subfloat[\label{fig:kotime_particle_s10}]{}
\subfloat[\label{fig:kotime_particle_s10_period}]{}
\includegraphics[scale=.85]{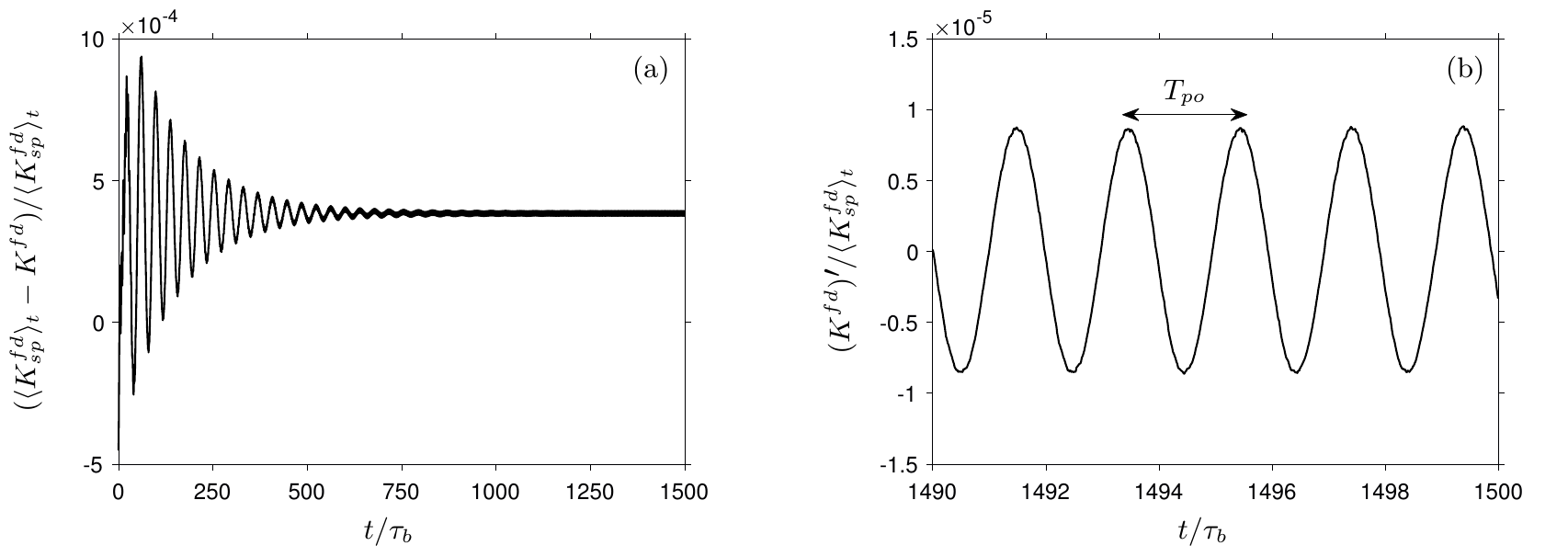}
\caption{(a) Time evolution of the
difference between the box-averaged kinetic energy in the single-phase case
($\mean{\kenfd_{sp}}_{t}$) and in run S10 ($\kenfd$). (b) Zoom of the
fluctuations of the box-averaged kinetic energy, $(\kenfd)^\prime$, around its
mean value in the periodic regime, shown for the final few cycles of case S10.}
\label{fig:kotime_s10}
\end{figure}

\begin{figure}[p]
\centering
\includegraphics[scale=1]{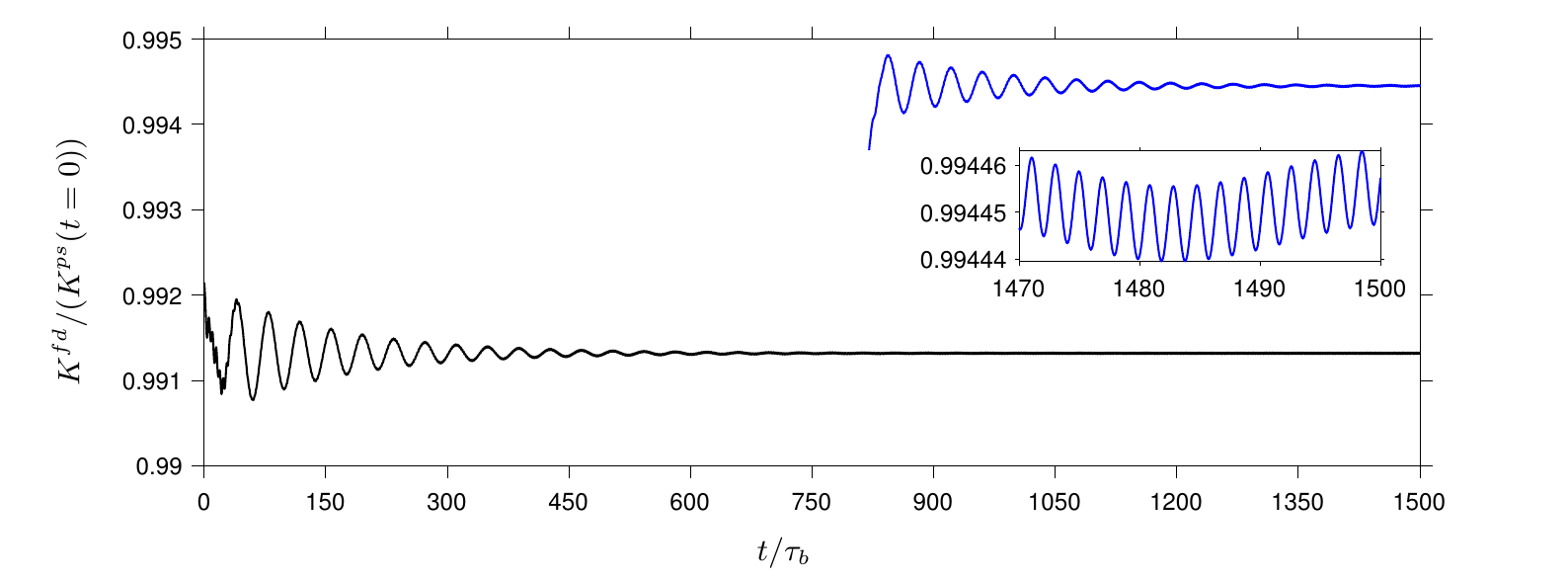}
\caption{Time evolution of the box-averaged kinetic energy of the fluctuations
$(\kenfd)$ in the baseline case S10 (\mbox{\protect\blackline}) and after grid
refinement, S10-F (\mbox{\protect\blueline}). The inset shows a zoom over the
last few cycles of the latter run.}
\label{fig:kotime_s10_fine}
\end{figure}

The initial condition for case S10-F was taken from the state of the system in
case S10 at $t=820.50\,\tscaleb$. The respective velocity field was first
interpolated on the finer grid and the simulation was subsequently continued
until $t=1500\,\tscaleb$. The interpolation introduced disturbances in the
former steady flow-field and gave rise to a new transient, as evidenced by the
discontinuity in the time evolution of box-averaged kinetic energy in case S10-F
shown in \cref{fig:kotime_s10_fine}. The transient behavior is, as before,
marked by a damped oscillation, which decays in time and slowly approaches a
periodic state. The time-average of $(\kenfd)$ differs by 0.3\% of
$\kenspec(t=0)$ between the computations on the two grids. Despite the lengthy
time integration, we observe that the well-defined periodic motion previously
seen in \cref{fig:kotime_particle_s10_period} is not yet fully established in
case S10-F (see the inset in \cref{fig:kotime_s10_fine}), whereas the signature
of the periodic motion is already clearly developed. Note that the remaining
slow modulation at the end of the simulation S10-F has an amplitude of
approximately 1.5 times the amplitude of the oscillations related to the
flow-through of the particle. The period of the latter oscillations in case
S10-F measures $T_{po}\approx1.94\,\tscaleb$ which is in close agreement with
case S10 (cf.\ \cref{fig:kotime_particle_s10_period}); for the amplitude of the
fluctuations we observe a discrepancy of approximately 7\% between the two
grids.

\begin{figure}[p]
\centering
\captionsetup[subfigure]{labelformat=empty}
\subfloat[\label{fig:spanwise_pos_s10fine}]{}
\subfloat[\label{fig:spanwise_vel_s10fine}]{}
\includegraphics[scale=.85]{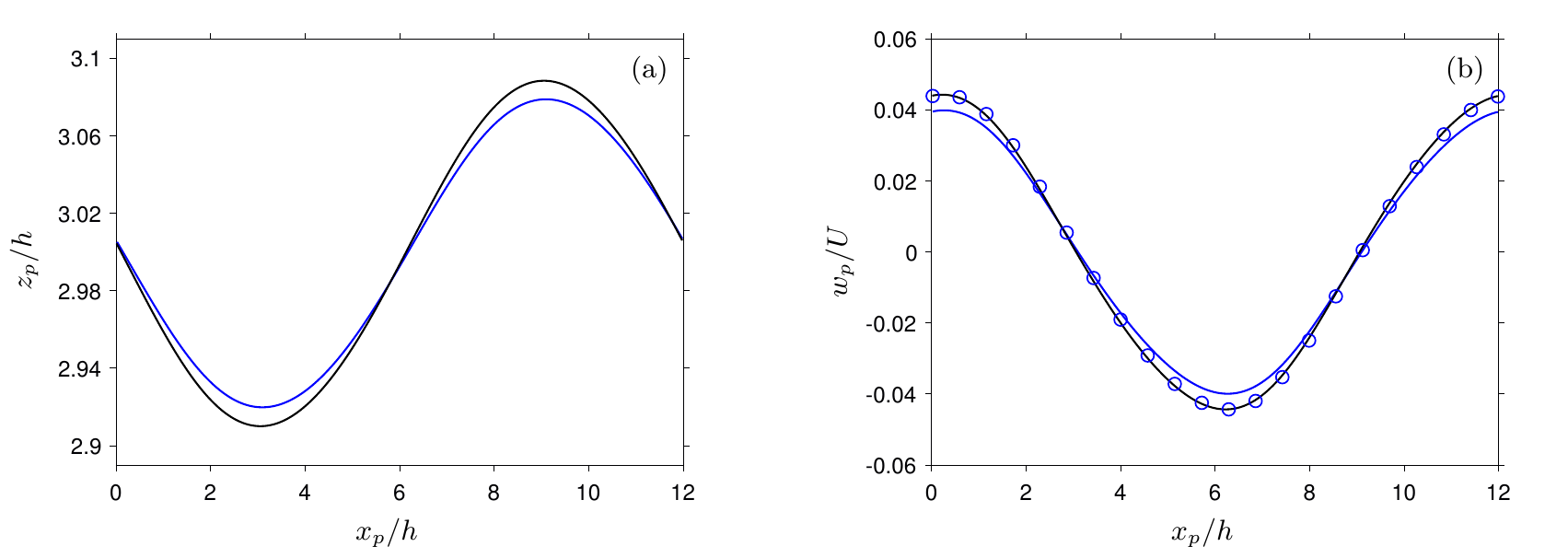}
\caption{Data over one cycle in the asymptotic regime in case S10
(\mbox{\protect\blackline}) and using the refined grid of S10-F
(\mbox{\protect\blueline}) showing: (a) the spanwise particle position; (b) the
spanwise particle velocity component. In (b), the blue open circles
(\mbox{\protect\oblue}) represent the data from case S10-F scaled by the ratio
of the two amplitudes. }
\label{fig:spanwise_pos_velo_s10fine}
\end{figure}

Concerning the spanwise particle motion in the asymptotic regime, we find a good
agreement between the results obtained with the two grids (S10 and S10-F) for
the particle position (\cref{fig:spanwise_pos_s10fine}) and velocity
(\cref{fig:spanwise_vel_s10fine}). For both signals, case S10-F exhibits
amplitudes that are lower by 11\% than in case S10. The shapes of the curves,
however, remain unaltered, and re-scaling the spanwise velocity of the finer
case recovers the signal of the coarser run. This confirms the negligible
phase-lag between the two curves in \cref{fig:spanwise_pos_velo_s10fine}.

\begin{table}[p]
\centering
\caption{Geometrical dimensions of the plane Couette numerical set-up used to
compute Nagata's equilibrium solutions. The half wall separation is given by $h$
and an illustration of the computational domain is shown in
\cref{fig:couette_geometry}.}
\begin{tabular}{ccccc}
\hline\hline
Domain ID & $\domain{x}/h$ & $\domain{z}/h$ & $\wllx/h$ & $\wllz/h$\tabularnewline\hline
$\gv{\Omega}_1$ & 6 & 3 & 6 & 3
\tabularnewline
$\gv{\Omega}_2$ & 12 & 6 & 12 & 6
\tabularnewline
\hline\hline
\end{tabular}
\label{tb:trackeqsol}
\end{table}

\begin{table}[p]
\centering
\caption{
Physical parameters of the stable equilibrium solution selected to study
the interaction between finite-size particles and exact coherent
structures.}
\begin{tabular}{cccccccc}
\hline\hline
Domain ID & $\reynolds$ & $\retau$ & $\dissipation$ & $\wllx/h$ & $\wllz/h$  &
$\domain{x}/\deltanu$ & $\domain{z}/\deltanu$
\tabularnewline\hline
$\gv{\Omega}_2$ & $132.25$ & $15.87$ & $1.90$ & $12$ & $6$ & $190.38$ & $95.19$
\tabularnewline
\hline\hline
\end{tabular}
\label{tb:nagatare132}
\end{table}

\begin{table}[p]
\centering
\caption{Physical parameters of the runs with Nagata's solution at $Re=132.25$
on the upper branch of $\gv{\Omega}_2$ seeded with single or multiple
finite-size particles. The diameter of the particles is $\dpart/h=0.156$
($\dplus=2.48$). The fluid phase is discretized with $768\times128\times384$
grid points and the particle resolution measures $\dpart/\Delta x=10$ in all
cases.}
\begin{tabular}{cccccccc}
\hline\hline
Run ID & \# Particles & $\rhop/\rhof$ & $\stokes$ & $\stokesplus$ &
$\galileo$ & $T_{obs}/\tscaleb$ & Colormap \tabularnewline\hline
S04 & $1$ & $4$ & $0.22$ &  $1.37$ &  $3.16$ & $110$
& \textcolor{cS04}\straightline
\tabularnewline
S06 & $1$ & $6$ & $0.33$ &  $ 2.05$ &  $4.08$ & $110$
& \textcolor{cS06}\straightline
\tabularnewline
S08 & $1$ & $8$ & $0.44$ &  $2.73$  &  $4.83$ & $110$
& \textcolor{cS08}\straightline
\tabularnewline
S10 & $1$ & $10$ & $0.56$ & $3.41$ & $5.48$ & $1500$
& \textcolor{cS10}\straightline
\tabularnewline
S12 & $1$ & $12$ & $0.67$ &  $4.10$  &  $6.06$ & $110$
& \textcolor{cS12}\straightline
\tabularnewline
S16 & $1$ & $16$ & $0.89$ &  $5.46$  &  $7.07$ & $110$
& \textcolor{cS16}\straightline
\tabularnewline
S20 & $1$ & $20$ & $1.11$ &  $6.83$  &  $7.96$ & $110$
& \textcolor{cS20}\straightline
\tabularnewline
S25 & $1$ & $25$ & $1.39$ &  $8.53$  &  $8.95$ & $110$
& \textcolor{cS25}\straightline
\tabularnewline
S30 & $1$ & $30$ & $1.67$ &  $10.24$ &  $9.84$ & $110$
& \textcolor{cS30}\straightline
\tabularnewline
S35 & $1$ & $35$ & $1.94$ &  $11.95$ &  $10.65$ & $110$
& \textcolor{cS35}\straightline
\tabularnewline
S40 & $1$ & $40$ & $2.22$ &  $13.66$ &  $11.41$ & $110$
& \textcolor{cS40}\straightline
\tabularnewline
S60 & $1$ & $60$ & $3.33$ &  $20.48$ &  $14.03$ & $110$
& \textcolor{cS60}\straightline
\tabularnewline
S80 & $1$ & $80$ & $4.44$ &  $27.31$ &  $16.23$ & $110$
& \textcolor{cS80}\straightline
\tabularnewline
M10 & $10$ & $10$  & $0.56$ & $3.41$ & $5.48$ & $120$
& N/A
\tabularnewline
C10\footnote {The particle motion is constrained in the z-direction}
 & $10$ & $10$  & $0.56$ & $3.41$ & $5.48$ & $116$
& N/A \tabularnewline
\hline\hline
\end{tabular}
\label{tb:particleparams}
\end{table}

\begin{table}[p]
\caption{Parameter points on the upper-branch of Nagata's solution
for which a stability analysis was performed.}
\centering
\begin{tabular}{ccc|ccc|ccc}
\hline\hline
\multicolumn{3}{c}{$\gv{\Omega}_1$: $(\wllx,\wllz)=(6h,3h)$}         &
\multicolumn{3}{c}{$\gv{\Omega}_2$: $(\wllx,\wllz)=(12h,6h)$}        &
\multicolumn{3}{c}{$\gv{\Omega}_3$: $(\wllx,\wllz)=(4\pi h,2\pi h)$}
\tabularnewline
\hline
$\dissipation$ & $Re$ & Stable? &
$\dissipation$ & $Re$ & Stable? &
$\dissipation$ & $Re$ & Stable?
\tabularnewline
\hline
$1.85$ & $171.46$ & No  & $1.86$ & $130.00$ & \textbf{Yes} & $1.83$ &
$130.05$ & \textbf{Yes}
\tabularnewline
$1.90$ & $172.28$ & No  & $1.89$ & $131.40$ & \textbf{Yes} & $1.86$ &
$131.04$ & \textbf{Yes}
\tabularnewline
$1.94$ & $173.58$ & No  & $1.90$ & $132.25$ & \textbf{Yes} & $1.88$ &
$132.25$ & \textbf{Yes}
\tabularnewline
$1.97$ & $174.72$ & No  & $1.92$ & $133.19$ & \textbf{Yes} & $1.90$ &
$133.38$ & \textbf{Yes}
\tabularnewline
$2.01$ & $176.84$ & No  & $1.93$ & $134.22$ & No           & $1.93$ &
$135.43$ & No
\tabularnewline
$2.04$ & $178.49$ & No  & $1.95$ & $135.33$ & No           & $1.94$ &
$136.59$ & No
\tabularnewline
$2.06$ & $180.31$ & No  & $1.96$ & $136.52$ & No           & $1.96$ &
$137.83$ & No
\tabularnewline
$2.19$ & $190.62$ & No  & $1.98$ & $137.78$ & No           & $1.97$ &
$139.15$ & No
\tabularnewline
$2.27$ & $199.32$ & No  &       &         &             & $1.99$ &
$140.54$ & No
\tabularnewline
$2.35$ & $209.56$ & No  &       &         &             &       &
        &
\tabularnewline
\hline\hline
\end{tabular}
\label{tb:stable_unstable}
\end{table}

 \end{document}